\documentclass[10pt,conference]{IEEEtran}

\usepackage{booktabs}
\usepackage{multirow}
\usepackage{siunitx}
\usepackage{amsmath}
\usepackage{amssymb}
\usepackage[linesnumbered,ruled,vlined]{algorithm2e}
\usepackage{listings}
\usepackage{hyperref}
\usepackage{caption}
\usepackage{listings}
\usepackage{stmaryrd}
\usepackage{graphicx}
\usepackage{subcaption}
\usepackage[most]{tcolorbox}
\usepackage{lipsum} 
\tcbuselibrary{listingsutf8}
\usepackage{cleveref}
\usepackage{dsfont} 
\usepackage{wasysym} 
\usepackage{import} 
\usepackage{caption} 
\usepackage{dsfont} 
\usepackage{soulutf8} 
\usepackage{makecell}
\usepackage{tabularx}
\usepackage{threeparttable}
\sethlcolor{YellowGreen}
\usepackage{algpseudocode}
\usepackage{tikz}
\usepackage{tabularx}
\usetikzlibrary{arrows.meta, positioning}
\usepackage{xcolor}
\definecolor{turquoise}{rgb}{0.28, 0.82, 0.8}
\definecolor{darkgreen}{RGB}{54, 156, 90}
\definecolor{RedOrange}{HTML}{ED135A}
\definecolor{YellowGreen}{HTML}{98CC70}
\definecolor{NavyBlue}{HTML}{006EB8}
\newcommand{\projName}{\textsc{SpecFix}\xspace}

\newtheorem{definition}{Definition}
\newtheorem{example}{Example}

\DeclareMathOperator*{\argmax}{arg\,max}

\newcommand{\changes}[1]{%
#1\xspace
}

\newcommand{\ie}{\textit{i}.\textit{e}.,}

\newcolumntype{T}{S[table-format=2.1, table-align-text-post=false]}

\definecolor{promptbg}{RGB}{255,255,255}      
\definecolor{prompttitle}{RGB}{255,255,255}

\newtcolorbox{promptbox}[1][]{
  enhanced,                      
  colback=promptbg,    
  colframe=NavyBlue,
  coltitle=prompttitle,
  boxrule=0.8pt,       
  arc=1mm,             
  left=2mm,            
  right=2mm,           
  top=1mm,             
  bottom=1mm,          
  width=\columnwidth,  
  fonttitle=\bfseries, 
  title=#1             
}

\newtcolorbox{responsebox}[1]{
    colback=green!5!white,
    colframe=green!65!black,
    fonttitle=\bfseries,
    title=#1,
    sharp corners,
    boxrule=1pt,
    width=\columnwidth,
}

\newenvironment{findingBox}[2]{%
	\begin{tcolorbox}[
colframe=black!80,
colback=gray!10,
 boxrule=.5pt,
 left=1pt,
 right = 1pt,
 top=0pt,
 bottom=0pt,
 size=small,
 fonttitle=\bfseries,
coltitle=black,
boxrule=0.4mm,
arc=2mm
 ]{\textbf{Observation #1: }#2} 
}{%
	\end{tcolorbox}
}

\newenvironment{rqBox}[2]{%
	\begin{tcolorbox}[
colframe=black!80,
colback=gray!10,
 boxrule=.5pt,
 left=1pt,
 right = 1pt,
 top=0pt,
 bottom=0pt,
 size=small,
 fonttitle=\bfseries,
coltitle=black,
boxrule=0.4mm,
arc=2mm
 ]{\textbf{RQ#1: }#2} 
}{%
	\end{tcolorbox}
}

\begin{document}

\title{\changes{Automated Repair of Ambiguous Problem Descriptions for LLM-Based Code Generation}}


\author{\IEEEauthorblockN{Haoxiang Jia\IEEEauthorrefmark{1}, Robbie Morris\IEEEauthorrefmark{2}, He Ye\IEEEauthorrefmark{2}, Federica Sarro\IEEEauthorrefmark{2} and Sergey Mechtaev\IEEEauthorrefmark{1}\IEEEauthorrefmark{3}}
\IEEEauthorblockA{\IEEEauthorrefmark{1}Peking University, Beijing, China\\
Emails: haoxiangjia@stu.pku.edu.cn, mechtaev@pku.edu.cn}
\IEEEauthorblockA{\IEEEauthorrefmark{2}University College London, United Kingdom\\
 Emails: \{robbie.morris.22,he.ye,f.sarro\}@ucl.ac.uk}
\thanks{\IEEEauthorrefmark{3}Sergey Mechtaev is the corresponding author.}}

\IEEEoverridecommandlockouts 

\maketitle

\begin{abstract}
The growing use of large language models (LLMs) has increased the importance of natural language (NL) in software engineering. However, ambiguity of NL can harm software quality, as unclear \changes{problem descriptions} may lead to incorrect program generation. Detecting and resolving such ambiguity is challenging, motivating our introduction of the automated repair of ambiguous NL \changes{descriptions}, which we approach by reducing code generation uncertainty and better aligning NL with input–output examples. Ambiguity repair is difficult for LLMs because they must understand how their interpretation of a \changes{description} changes when the text is altered. We find that directly prompting LLMs to clarify ambiguity often produces irrelevant or inconsistent edits. To address this, we decompose this task into two simpler steps: (1) analyzing and repairing the LLM’s interpretation of the \changes{description} --- captured by the distribution of programs it induces --- using traditional testing and program repair, and (2) refining the \changes{description} based on distribution changes via a method we call contrastive specification inference. We implement this approach in a tool called \projName and evaluate it using four state-of-the-art LLMs (GPT‑4o, \changes{GPT‑4o‑mini}, DeepSeek‑V3, and Qwen2.5‑Coder‑32B‑Instruct) on three popular code generation benchmarks (HumanEval+, MBPP+ and \changes{LiveCodeBench}). Without human intervention or external information, \projName modified 43.58\% of \changes{descriptions}, improving Pass@1 on the modified set by 30.9\%. This yields a 4.09\% absolute improvement across the entire benchmark. Repairs also transfer across models: \changes{descriptions} repaired for one model improve other models’ performance by 10.48\%.
\end{abstract}

\section{Introduction}
\label{sec:introduction}

\begin{figure*}
    \centering
    \includegraphics[width=0.95\textwidth]{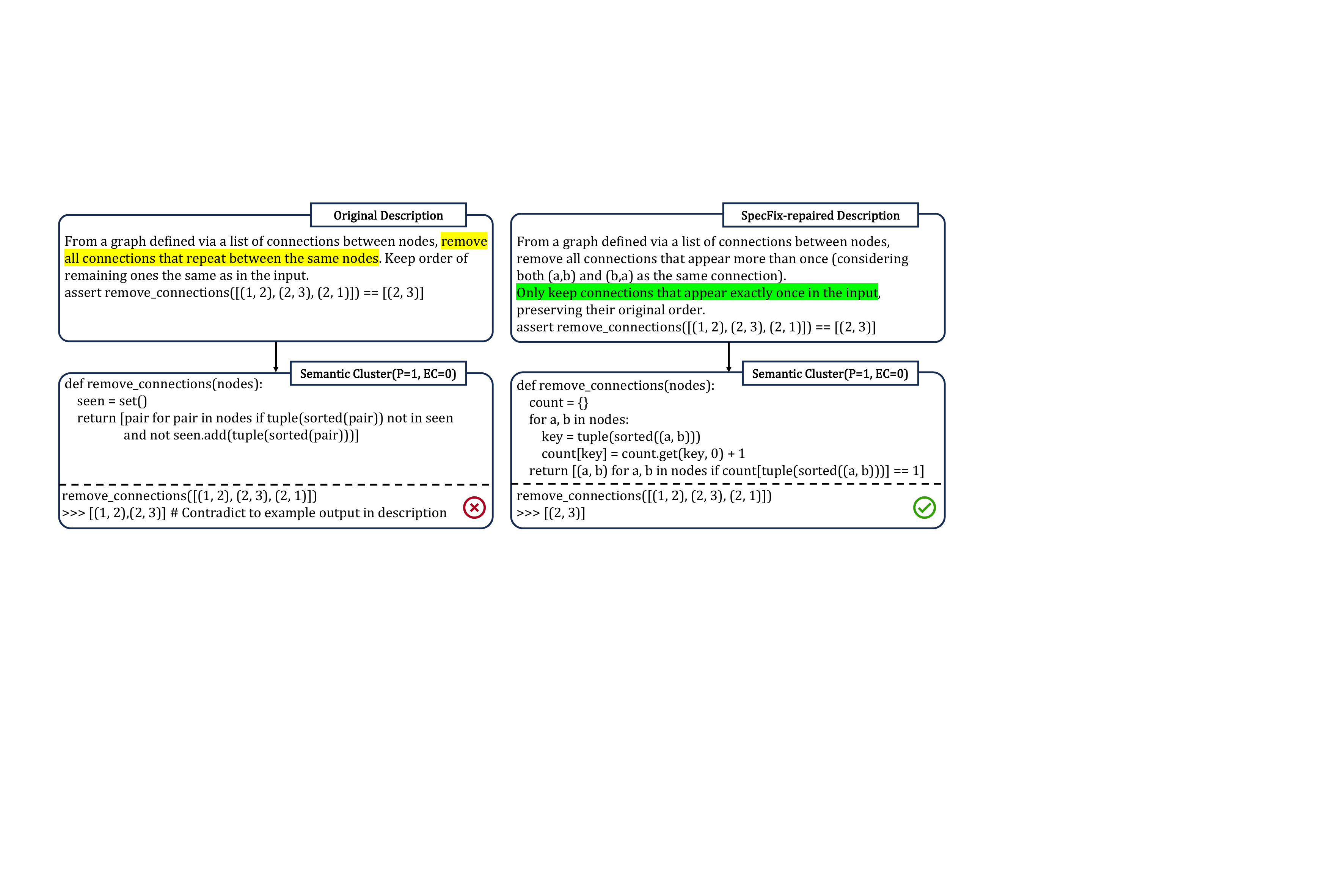}
    \caption{The requirement to delete ``repeating'' connections, highlighted with \colorbox{yellow}{\phantom{x}}, is ambiguous: it can be interpreted as either (1) deleting all connections that appear more than once, or (2) deleting all occurrences starting from the second one. Despite the presence of a clarifying example supporting the first interpretation, DeepSeek-V3 consistently outputs the second interpretation. A SpecFix-generated repair, highlighted with \colorbox{green}{\phantom{x}}, enables a consistent generation of the interpretation conforming to the example. 
    }
    \label{fig:motivating-example1}
    \vspace{-3mm}
\end{figure*}

Natural language (NL) has assumed an important role in software development due to the powerful capabilities of large language models (LLMs). LLMs use NL inputs, such as prompts or chat-based interactions, as a form of software requirements to generate code. However, the inherent ambiguity and susceptibility to misinterpretation in NL can impact the quality of generated code --- introducing bugs or resulting in implementations that deviate from the intended purpose. For example, Vijayvargiya et al.~\cite{vijayvargiya2025interactive} observed that the use of ambiguous task descriptions reduced the performance of models by 20\%. Previous work~\cite{mu2023clarifygpt} analyzed semantic discrepancies in programs sampled from an LLM to quantify the \changes{description's} ambiguity, and then prompted the LLM to ask user clarifying questions. However, our experiments demonstrate that LLMs' clarifying questions are often redundant, i.e. irrelevant to LLMs' own interpretation of the problem, leading to verbose and ineffective \changes{descriptions} and imposing an additional cognitive burden on the users.

To overcome the above limitations, this paper introduces the problem of \emph{automated repair of ambiguous natural language \changes{problem descriptions}}, defining it as the task of systematically modifying \changes{descriptions} to ensure adherence to predefined quality criteria. Although in some cases human input is required to fix ambiguity, our insight is that many ambiguities can be resolved fully-automatically based on two key observations. First, we can modify \changes{problem descriptions} to reduce LLM's code generation uncertainty by increasing the probability that the LLM generates the most likely interpretation of ambiguous language. Second, while humans often use input-output examples to clarify their intent~\cite{le2014flashextract}, state‐of‐the‐art (SOTA) LLMs struggle to interpret these examples in the presence of ambiguous language. In this context, we can automatically remove the ambiguity by aligning the natural language with the clarifying examples to facilitate the model's correct interpretation.

Our experience shows that SOTA LLMs struggle with directly repairing ambiguous \changes{problem descriptions}, often producing irrelevant or inconsistent changes. \changes{We hypothesize that this is because for an LLM it is hard to reflect on how changes to natural language text affect the programs it would generate based on the altered text.} To overcome this problem, we propose \projName, an  approach to repair \changes{problem description} that addresses this challenge by decomposing the task into simpler subtasks: it first \emph{repairs the distribution of programs} this \changes{description} induces and then maps back the change to the \changes{description} via \emph{contrastive specification inference}. 

When repairing the program distribution, \projName uses two quality criteria. To lessen LLM's code generation uncertainty, \projName aims to reduce \emph{semantic entropy}~\cite{kuhn2023semantic}, which is computed on the distribution of equivalence classes of programs clustered based on their input-output behavior, since high semantic entropy implies a high number of different interpretations. To align \changes{a problem description} with clarifying examples, \projName aims to increase \emph{example consistency}, our novel measure that indicates to what degree programs in the distribution satisfy the examples. When all sampled programs exhibit behavior contradictory to the examples, \projName applies automated program repair~\cite{le2019automated} to fix programs in the distribution. The natural language \changes{description} is then iteratively improved by \projName's contrastive specification inference prompt, which minimally modifies the text to prioritize desirable interpretations over undesirable ones.

We conducted a comprehensive evaluation of SpecFix using \changes{problem descriptions} from there widely used code generation benchmarks: HumanEval+, MBPP+ and \changes{LiveCodeBench}, and four SOTA LLMs: GPT-4o, \changes{GPT-4o-mini}, DeepSeek-V3 and Qwen2.5-Coder-32b-Instruct. Our results reveal that solely based on the examples embedded within these \changes{descriptions} --- without relying on any external information or user feedback --- SpecFix modifies 43.58\% of the \changes{descriptions}, leading to a 30.9\% improvement in model Pass@1 on the modified \changes{descriptions}. Across the entire benchmark, this corresponds to an absolute increase of 4.09\% in overall Pass@1. Importantly, we demonstrate cross-model generalizability: repairs generated by one model boost the performance of other models by 10.48\%. Finally, we found that the \projName’s repairs result in only modest increases in \changes{description} length.

In summary, the paper makes the following contributions:

\begin{itemize}
\item The first approach to automated repair of \changes{natural language problem descriptions for LLM-based code generation}.
\item A novel \changes{problem description} quality criterion, example-consistency, motivated by LLM's inability to extract the latent intent of clarifying input-output examples.
\item A repair mechanism via reflection on the problem distribution using \changes{contrastive specification inference}.
\item An extensive evaluation showing the effectiveness of \changes{description} repairs, and their cross-model generalizability.
\end{itemize}

All code, scripts, and data necessary to reproduce this work are available at \url{https://github.com/msv-lab/SpecFix}.


\section{Motivating Examples}
\label{sec:motivating}

\begin{figure*}
    \centering
    \includegraphics[width=\textwidth]{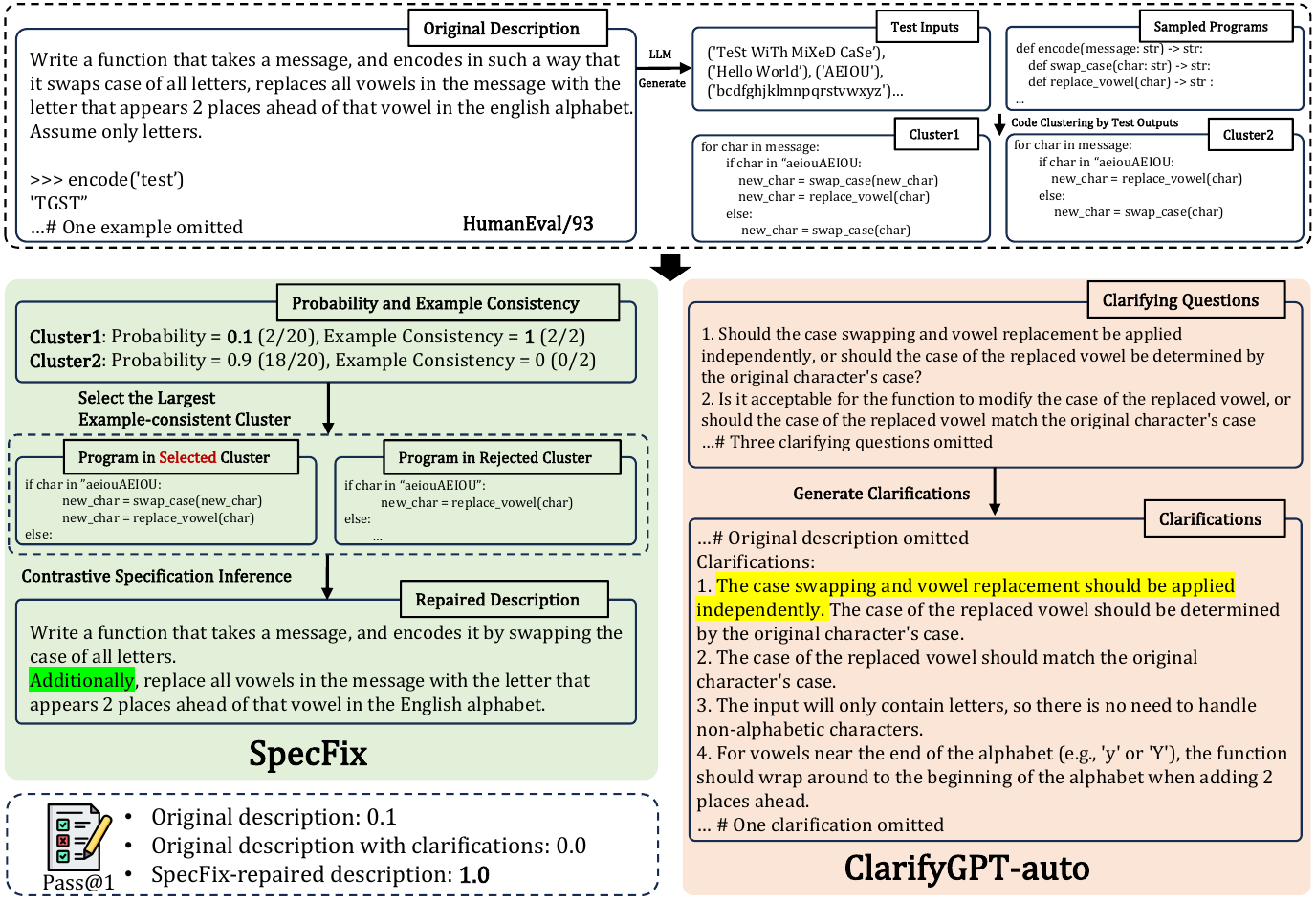}
    \caption{A straightforward application of an LLM to repair ambiguity with ClarifyGPT-auto results in inconclusive (marked with \colorbox{yellow}{\phantom{x}}) or irrelevant clarifications. \projName decomposes the problem into \emph{program distribution repair} and \emph{contrastive specification inference} to produce a correct, minimal disambiguation highlighted with \colorbox{green}{\phantom{x}}. The responses are generated with DeepSeek-V3.}
    \label{fig:motivating-example2}
    \vspace{-3mm}
\end{figure*}

In this section, we present three motivating examples to illustrate problems of ambiguous \changes{problem descriptions}: (1) LLMs' inability to extract the latent intent of input-output examples to clarify ambiguous language, (2) \changes{limitations of existing LLM-based approaches that prevent them from rectifying ambiguity}, and (3) the distinction between \changes{problem description} repair and reasoning methods.

To characterize an LLM’s interpretation of \changes{a problem description}, we analyze the distribution of programs this description induces. Following previous work~\cite{fan2024oracle,mu2023clarifygpt}, we compute this interpretation by sampling programs and partitioning them into equivalence classes based on their input-output behavior. We refer to these classes as \emph{semantic clusters}.

\subsection{Natural Language Ambiguity Degrades LLM Performance}
\label{sec:ambiguity_effect}
Consider the \changes{problem description} in \Cref{fig:motivating-example1}, which specifies to remove all edges that repeat between nodes in a graph. This natural language description is ambiguous, admitting two interpretations: (1) delete all connections that appear more than once, or (2) delete all occurrences starting from the second one. 
Although the example provided with this \changes{description} clearly illustrates the first interpretation, when we sampled 20 programs generated by DeepSeek-V3~\cite{liu2024deepseek} for this \changes{description}, all the programs fell into a single semantic cluster (i.e., they exhibited the same behavior) aligning with the second interpretation, which is wrong. One of such programs is shown at the bottom left of \Cref{fig:motivating-example1} as representative of the 20 programs. Despite matching one another ($P=1$), every program failed the embedded example, consistently adopting the second interpretation and thus disregarding the clarifying I/O example.
We measure this phenomenon using example consistency (EC)—the fraction of I/O examples that programs in a cluster satisfy. Here, the cluster’s EC is $0$, indicating a complete mismatch between the model’s behavior and the intended semantics.

This example demonstrates that subtle ambiguity can dominate the model’s performance, which was also observed in previous work~\cite{vijayvargiya2025interactive}. Our solution to the automated repair problem, \projName, resolved the above ambiguity by explicitly stating: ``Only keep connections that appear exactly once in the input.'' Under this disambiguated specification, DeepSeek-V3 generates the correct implementation with $P=1$.

\begin{findingBox}{1}{
Subtle ambiguities in natural language \changes{problem descriptions} can lead LLMs to generate incorrect code, even when clarifying I/O examples are present. Aligning the natural language description with the examples enables automatic, precise disambiguation.}
\end{findingBox}

\begin{figure*}
    \centering
    \includegraphics[width=\textwidth]{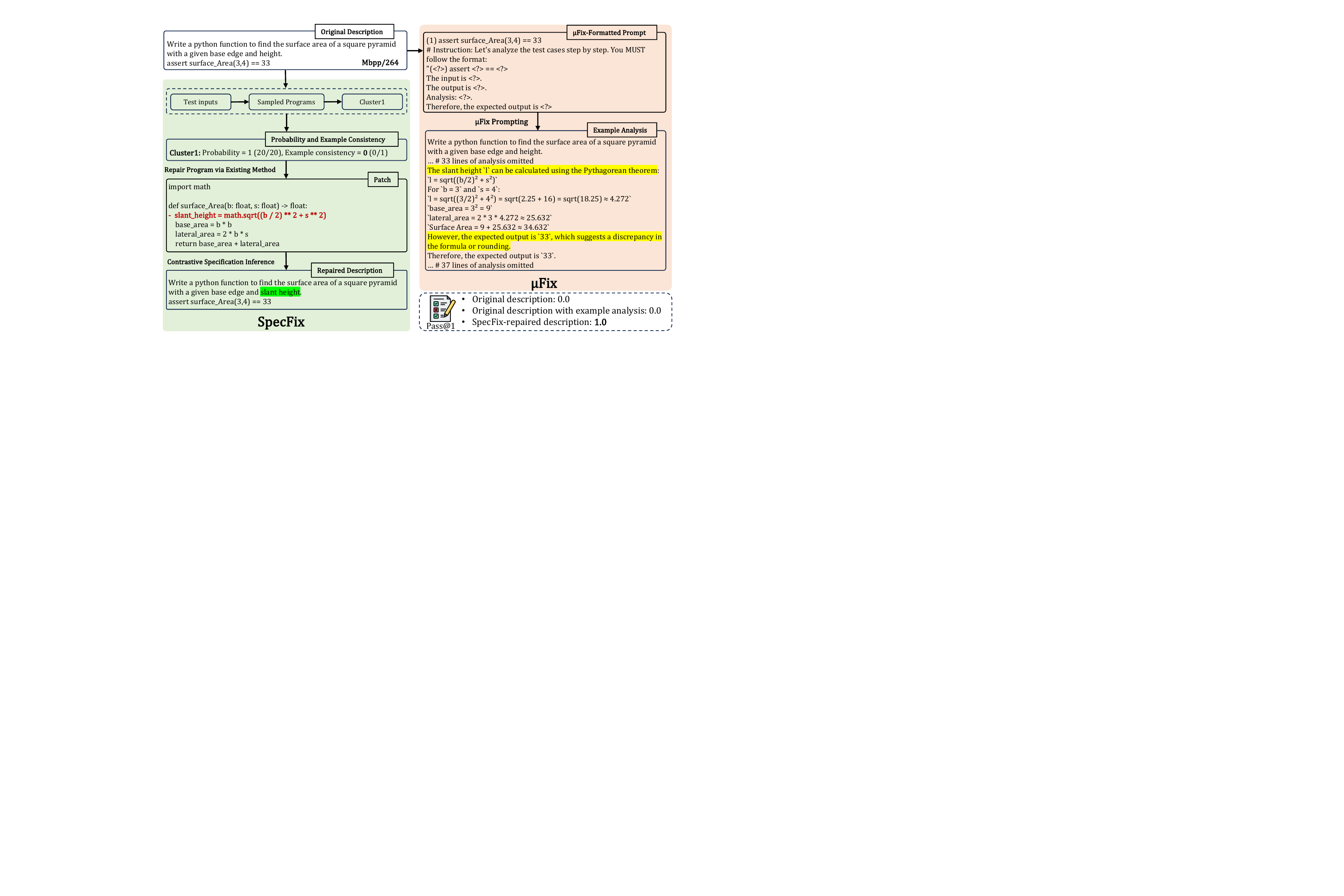}
    \caption{\projName repairs the \changes{description} ambiguity by first using the example to repair programs in the distribution, and then applying contrastive specification inference on the programs before and after repair; $\mu$Fix fails to correctly analyze examples. \colorbox{yellow}{\phantom{x}} marks the incorrect example analysis, while \colorbox{green}{\phantom{x}} marks the correct disambiguation. This example uses DeepSeek-V3.}
    \label{fig:motivating-example3}
\end{figure*}

\subsection{\changes{Limitations of Existing LLM-Based Approaches}}

LLMs are obvious candidates for addressing the ambiguity repair problem. One prominent approach is ClarifyGPT~\cite{mu2023clarifygpt}, which marks \changes{a problem description} as ambiguous whenever an LLM's interpretation of this \changes{description} contains at least two semantic clusters of programs, and asks clarifying questions to differentiate programs from these clusters. Although this approach was not designed for automated repair, it can be adapted for repair by using its ``user simulation prompt'' to answer the generated clarifying questions based on the examples embedded in the \changes{problem description}, and adding the resulting clarifications to the original \changes{description}. We refer to this approach as ClarifyGPT-auto.

\Cref{fig:motivating-example2} shows that the lack of a conjunction such as ``and'', ``or'' or ``then'' in the sentence ``swap cases of all letters, replace all vowels'' (top left) confuses DeepSeek-V3. It generates the correct interpretation, first swap cases of all letters and then replace vowels, only in 10\% of samples, corresponding to the Cluster 1, and in 90\% of samples swaps case of only consonants, despite the fact that the clarifying example clearly indicates the first interpretation. When applying ClarifyGPT-auto to repair this ambiguity, it asks five questions, four of which are redundant, as they do not lead to alternative interpretations, and answers the only relevant question about operation dependency inconclusively. Consequently, the pass rate for the \changes{problem description} after clarification drops to 0. 

This example shows the challenge of reasoning about ambiguity. To overcome it, \projName first analyses and repairs LLM's interpretation of \changes{a problem description} embodied in the distribution of semantic clusters. It measures the probability ($P$) and example consistency ($\mathrm{EC}$) for each cluster, and based on that repairs the distribution. In this example, the repair aims to increase the probability of the largest cluster that is fully consistent with the examples ($\mathrm{EC}=1$). Second, \projName employs \emph{contrastive specification inference} by prompting the LLM to generate the minimal revision to the \changes{problem description} which enables the selected cluster and disables the rejected cluster. This process results in a repair that adds a simple clarification ``Additionally'', showing that the two operations are sequential. Using the repaired \changes{description}, DeepSeek-V3 generates a correct program in 100\% of cases.

\begin{findingBox}{2}{LLMs struggle with ambiguity repair, which results in redundant or misleading clarifications. By decomposing the task into simpler subtasks, \projName achieves small, intent‑aligned \changes{problem description} repairs.}
\end{findingBox}

\subsection{\changes{Problem Description} Repair vs. Reasoning}

Some issues described above can be addressed by enhancing LLMs' reasoning capabilities. For example, $\mu$Fix, a state-of-the-art reasoning approach for code generation ``repairs'' its reasoning by analyzing why generated code fails provided examples~\cite{tian2025fixing}. Although \changes{problem description} repair and reasoning are complementary problems, we observe $\mu$Fix also struggles to reason correctly in the presence of natural language ambiguity.

\Cref{fig:motivating-example3} shows an ambiguous \changes{problem description} that does not specify if the input height is vertical height or slant height of a pyramid. Although it can be inferred from the clarifying example that the input is slant height, it is non-trivial as it requires ``reverse engineering'' various versions of the algorithm. To reason about these \changes{descriptions}, $\mu$Fix first constructs a structured analysis prompt, intended to elicit a step‑by‑step derivation of how to transform the given test inputs into the expected outputs. If the model fails to generate correct programs using the original \changes{description} and the analysis, $\mu$Fix revises its analysis by summarizing incorrect code and derives an updated \changes{description}. In this example, it correctly identifies that the total surface area equals the base area plus the lateral surface. However, when the computed results diverge slightly from the target values for the provide IO example, $\mu$Fix mistakenly attributes the discrepancy to ``rounding'' rather than questioning its understanding of the height parameter. As a result, the generated programs pass the tests in 0\% of cases.

\projName addresses this limitation by repairing the program distribution. Specifically, sampling programs from DeepSeek-V3 and partitioning them into semantic clusters results in a single cluster with the example consistency 0, i.e. failing the example test. Inspired by Fan et al.~\cite{fan2023automated}, \projName applies program repair to fix a program from this cluster using the example as the correctness criteria. Next, \projName employs contrastive specification inference to concisely modify \changes{the problem description} so that it mirrors the behavior of the repaired program while diverging from the original program. Here, the repaired \changes{description} explicitly states that the height is the slant height. With the repaired \changes{description}, DeepSeek-V3 generates a correct program in 100\% of cases.

\begin{findingBox}{3}{The SOTA approach for reasoning about \changes{problem descriptions} struggles in the presence of natural language ambiguity, as it is hard to ``reverse engineer'' the clarifying example for an ambiguously defined problem. \projName resolves by repairing programs in the distribution so that they align with the clarifying examples.}
\end{findingBox}

\section{Background and Notation}
\label{sec:background}

A \emph{specification} (or \emph{requirement}) is defined as ``the authoritative description of the behaviors, properties, or results that the automatically‐generated program must satisfy''~\cite{srivastava2013template}. In practice, specifications may be expressed via formal logics, natural‐language docstrings, or sets of input/output examples. In this work we focus exclusively on functional specifications (\ie~observable program behavior) \changes{for LLM-based code generation prompts}, abstracting away from non‐functional concerns such as performance, resource usage, or coding style. Henceforth we use the term ``\changes{problem descriptions}''.

Let $\mathcal{D}$ denote the space of problem descriptions and $\mathcal{P}$ the set of all programs in a chosen programming language. An LLM can be formalized as a conditional probability distribution that, for given \changes{problem description} $D\in\mathcal{D}$, assigns probabilities to various programs $m(\,\cdot\mid D) : \mathcal{P} \to [0,1].$ $\mathrm{supp}(m(\,\cdot \mid y))$ is the support of this distribution, i.e. the set of values with non-zero conditional probability.

In practice, this distribution is not given explicitly, but it can be approximated by sampling $N$ programs $\{p_i\}_{i=1}^N\;\overset{\mathrm{i.i.d.}}{\sim}\;m(\cdot\mid D)$ and using their frequencies to approximate probabilities. We denote such an approximated distribution as $\hat{m}$.

To factor out superficial syntactic variation, we use semantic equivalence relation $\equiv$ on $\mathcal{P}$: programs $P$ and $Q$ are equivalent if they produce identical outputs on all valid inputs.

\begin{definition}[Semantic Cluster]
  Let $\mathcal{P}$ be a set of programs sampled from an LLM. Semantic clusters are equivalence classes of the sampled programs denoted as $\mathcal{P}/\!\equiv$, where for each $P\in\mathcal{P}$ its semantic cluster is denoted as $$[P]\triangleq\{\,Q\in\mathcal{P}\mid Q\equiv P\}.$$
\end{definition}

Since the LLM induces a distribution over individual programs, it also induces a distribution over equivalence clusters:
\[
m_{\equiv}\bigl([P]\mid D\bigr)\triangleq\sum_{Q\in[P]} m(Q\mid D).
\]

\changes{
\begin{example}\label{ex:clusters}
  In \Cref{fig:motivating-example2}, there are two semantic clusters: one, $[P_1]$, contains programs that first swap cases of all letters, and the second, $[P_2]$, contains programs that swaps case of only consonants. Based on the generated sample, the estimated distribution of equivalence classes is $$m_{\equiv}\bigl([P_1]\mid D\bigr) = 0.1\quad m_{\equiv}\bigl([P_2]\mid D\bigr) = 0.9$$
\end{example}}

A specification is \emph{ambiguous} if it admits multiple plausible semantic interpretations~\cite{gleich2010ambiguity}. Previous work demonstrated that task ambiguity increases model's uncertainty~\cite{hou2023decomposing}. A popular measure of uncertainty is the \emph{semantic entropy}, which is the entropy over the meaning-distribution:

\begin{definition}[Semantic Entropy~\cite{kuhn2023semantic}]\label{def:semantic_entropy}
  Let $m$ be an LLM, $\equiv$ be a semantic equivalence relation over responses, $m_\equiv$ be the corresponding conditional distribution over equivalence classes. The semantic entropy is defined as
\begin{equation*}
  \mathrm{SE}(m_\equiv(\,\cdot\mid x))\triangleq-\sum_y m_\equiv(y\mid x)\log m_\equiv(y\mid x).
\end{equation*}
\end{definition}

ClarifyGPT~\cite{mu2023clarifygpt} proposed to classify \changes{problem descriptions} as ambiguous if their semantic entropy is above zero, as it means having more than one semantic clusters of programs:
\begin{equation*}
  \mathrm{ambiguous}_\mathrm{ClarifyGPT}(D)\;\triangleq\;\mathrm{SE}(m_\equiv(\,\cdot\mid R))>0.
\end{equation*}
\changes{
\begin{example}
  For the distribution in \Cref{ex:clusters}, the semantic entropy is $- (0.1 \log(0.1) + 0.9 \log(0.9)) = 0.469.$ Since it is positive, ClarifyGPT classifies the description as ambiguous.
\end{example}}

\section{\projName}
\label{sec:methodology}

\projName is a framework for automatically repairing ambiguous NL \changes{problem descriptions} to enhance code generation performance. Inspired by ClarifyGPT's ambiguity measure, \projName aims to eliminate code generation uncertainty by reducing the semantic entropy $\mathrm{SE}$. Motivated by our observation that ambiguous language may result in the generation of programs inconsistent with input-output examples embedded in the \changes{description}, which may encode latent intents not explicitly stated in the text (\Cref{sec:ambiguity_effect}), \projName also maximizes alignment with example via a new measure of \emph{example consistency} $\mathrm{EC}$:

\begin{definition}[Example Consistency]\label{def:example_consistency}
  Let $m$ be an LLM, $D$ be \changes{a problem description} with embedded input-output examples $\{(x_i,y_i)\}_{i=1}^m$, and $m_\equiv(\,\cdot\mid D)$ be the corresponding distribution of semantic clusters. For a program $P\in\mathrm{supp}(m_\equiv(\,\cdot\mid D))$, the example consistency is defined as 
\begin{equation*}
      \mathrm{EC}(P,\,\{(x_i,y_i)\}_{i=1}^m) \;=\; \frac{1}{m}\sum_{i=1}^m \mathbb{I}\bigl(P(x_i)=y_i\bigr)
\end{equation*}
where \( \mathbb{I}(\cdot) \) is the indicator function. For a distribution of clusters $m_\equiv(\,\cdot\mid D)$, the example consistency is defined as
\begin{align*}
  &\mathrm{EC}(m_\equiv(\,\cdot\mid D),\,\{(x_i,y_i)\}_{i=1}^m) \;=\\
  &\quad\sum_{[P]\in m_\equiv(\cdot\mid D)} m_\equiv([P]\mid D)\, \mathrm{EC}(P,\,\{(x_i,y_i)\}_{i=1}^m)
\end{align*}
\end{definition}

\changes{\begin{example}
  Consider clusters in \Cref{fig:motivating-example2}, for which we defined the distribution in \Cref{ex:clusters}. Programs in $[P_1]$ pass 2 out of 2 tests, while those in $[P_2]$ pass 0 out of 2 tests. Thus, the example consistency of this distribution is $\mathrm{EC} = 0.1 * (2 / 2) + 0.9 * (0 / 2) = 0.1$.
\end{example}}

Given the above measures, we formally define \changes{problem description} repair as follows:

\begin{definition}[\changes{Problem Description} Repair]
  Let $D$ be \changes{a problem description} with embedded input-output examples $\mathrm{EX}$, $m$ be an LLM. The goal of automated problem description repair is to find a \changes{description} $D'$ such that
  \begin{align*}
    &\mathrm{SE}(m_\equiv(\,\cdot\mid D')) < \mathrm{SE}(m_\equiv(\,\cdot\mid D))\\
    \wedge\ &\mathrm{EC}(m_\equiv(\,\cdot\mid D'),\,\mathrm{EX}) > \mathrm{EC}(m_\equiv(\,\cdot\mid D),\,\mathrm{EX})
  \end{align*}
and the difference between \changes{descriptions} $\mathrm{diff}(D, D') < \tau$ for some predefined threshold $\tau$. For \changes{descriptions} without embedded input-output examples, the $\mathrm{EC}$ is omitted.
\end{definition}

The ultimate goal is to find \changes{a description} with $\mathrm{SE}=1$ --- meaning the LLM interpretation forms a single semantic cluster --- and $\mathrm{EC}=0$ --- meaning all generated programs pass the embedded examples. Since achieving such a perfect repair may not always be possible, we instead aim to improve these measures within k iterations. We call SpecFix’s minimally edited output a \emph{repaired \changes{description}}, since improving these measures helps to resolve ambiguity.


\projName's key innovation is in decomposing this task into simpler subtasks. First, it analyzes and repairs the distribution of programs the \changes{problem description} induces. Second, it maps the change to the distribution back into \changes{the problem description} via contrastive specification interference.

\subsection{Distribution Repair}

\Cref{alg:overall_algorithm} summarizes the workflow of \projName. It accepts four arguments: \changes{a problem description} $D$, an LLM $m$, the iteration bound $K$, and the sample size $N$. It first starts with extracting examples from \changes{problem descriptions} using the function \texttt{Extract\_Examples}, which we implement by prompting the same LLM $m$.

A key part of this algorithm is repairing the distribution of programs that $D$ induces, as this distribution characterizes LLM's interpretation of the \changes{problem description}. The algorithm approximates the distribution $\hat{m}_\equiv(\,\cdot\mid D)$ by sampling $N$ programs and partitioning them into semantic clusters using the function \texttt{Interpret}. In the body of this function, the partitioning algorithm is encapsulated by the functions \texttt{Generate\_Inputs} and \texttt{Partition}. In our implementation, we realize them by prompting the same LLM to generate test inputs $\{t_i\}_{i=1}^{j}$ with the aim of covering all functionality. Then, we partition programs based on the outputs they produce on these inputs --- programs with identical outputs are grouped into the same cluster, indicating equivalent behavior. Alternative implementations such as differential fuzzing~\cite{petsios2017nezha} or symbolic execution~\cite{sharma2025assessing} can also be applied in this context.

\begin{algorithm}[t]
\SetKwInOut{Input}{Input}
\SetKwInOut{Output}{Output}
\SetKwFunction{QuantifyReq}{Interpret}
\SetKwFunction{Entropy}{Entropy}
\SetKwFunction{ExampleConsistency}{ExampleCons}
\SetKwFunction{SpecFix}{SpecFix}
\SetKwFunction{OutputCheck}{Partition}
\SetKwFunction{ExampleExtractPrompt}{Extract\_Examples}
\SetKwFunction{TestGenerationPrompt}{Generate\_Inputs}
\SetKwFunction{ProbRepair}{Prob\_Based\_Repair}
\SetKwFunction{PatchRepair}{Patch\_Based\_Repair}
\SetKwFunction{ContrastiveInference}{Contrast\_Infer}
\SetKwFunction{ProgramRepair}{Program\_Repair}

\Input{\changes{Problem description} $D$, LLM $m$, iteration bound $K$, sample size $N$}
\Output{Repaired \changes{problem description} $D'$}

\BlankLine
\SetKwProg{Fn}{Function}{:}{}

\Fn{\QuantifyReq{$m$, $D$, $N$}}{
$\{p_i\}_{i=1}^{N} \sim m(\,\cdot\mid D)$\;
$\{t_i\}_{i=1}^{j} \leftarrow \TestGenerationPrompt(D)$\;
$\hat{m}_\equiv(\,\cdot\mid D) \;\leftarrow\; \OutputCheck \!\bigl(\{p_i\}_{i=1}^{N},\,\{t_i\}_{i=1}^{j}\bigr)$\;
\Return $\hat{m}_\equiv(\,\cdot\mid D)$
}

\Fn{\SpecFix{$D, m, K, N$}}{
  $\{(x_i,y_i)\}_{i=1}^{l} \leftarrow \ExampleExtractPrompt(D)$\; 
  $\hat{m}_\equiv(\,\cdot\mid D) \leftarrow \QuantifyReq(m, D, N)$\;
  $\mathrm{se} \leftarrow \mathrm{SE}(\hat{m}_\equiv(\,\cdot\mid D))$\;
  $\mathrm{ec} \leftarrow \mathrm{EC}(\hat{m}_\equiv(\,\cdot\mid D), \{(x_i,y_i)\}_{i=1}^{l})$\;
  \For{$i \leftarrow 1$ \KwTo $K$}{
    \uIf{$\mathrm{se} = 0 \,\land\, \mathrm{ec} = 1$}{
      \Return $D$  \tcp*{D is unambiguous}
    }
    $C\leftarrow \mathrm{supp}(\hat{m}_\equiv(\,\cdot\mid D))$\;
    \uIf{$\exists\,c\in C\text{ s.t. }\mathrm{EC}(c)=1$}{
      $[P_\mathrm{select}] \leftarrow \argmax_{c\in C\text{ s.t. }\mathrm{EC}(c)=1} |c|$\;
      $P_\mathrm{reject} \leftarrow \cup(C \setminus [P_\mathrm{select}])$\;
    }
    \Else{
      $[P_\mathrm{reject}] \leftarrow \argmax_{c\in C} |c|$\;
      $P_\mathrm{select} \leftarrow \ProgramRepair(P_\mathrm{reject}, \{(x_i,y_i)\}_{i=1}^{l})$
   }
    $D' \leftarrow \ContrastiveInference(D, P_\mathrm{select}, P_\mathrm{reject})$\;
    $\hat{m}_\equiv(\,\cdot\mid D') \leftarrow \QuantifyReq(m, D', N)$\;
    $\mathrm{se}' \leftarrow \mathrm{SE}(\hat{m}_\equiv(\,\cdot\mid D'))$\;
    $\mathrm{ec}' \leftarrow \mathrm{EC}(\hat{m}_\equiv(\,\cdot\mid D'), \{(x_i,y_i)\}_{i=1}^{l})$\;
    \uIf{$\mathrm{ec}' > \mathrm{ec} \,\land\, \mathrm{se}' < \mathrm{se}$}{
      $D \leftarrow D', \mathrm{ec} \leftarrow \mathrm{ec}', \mathrm{se} \leftarrow \mathrm{se}'$\;
    }
  }
  \Return $D$\;
}
\caption{\projName Algorithm}
\label{alg:overall_algorithm}
\end{algorithm}


To resolve ambiguities, \projName employs two repair strategies: probability‐guided repair and program repair based \changes{problem description} repair. If one or more clusters pass all examples ($\mathrm{EC}=1$), \projName invokes the probability‐guided repair strategy; otherwise --- when no cluster can pass all examples --- it resorts to the program repair based \changes{description} repair.


\paragraph*{Probability-guided Repair}

To reduce semantic entropy, \projName prioritizes the most likely interpretation. First, it prioritizes semantic clusters that pass all examples ($\mathrm{EC}=1$) over other clusters. The process is described in Algorithm~\ref{alg:overall_algorithm} in lines 15--17, \changes{and is illustrated in \Cref{fig:motivating-example2}, where the low-probability cluster (0.1) is prioritized over the high-probability cluster (0.9), because the former has higher example consistency.} If there are multiple clusters with $\mathrm{EC}=1$, the algorithm selects one with the highest probability, treating all others as rejected clusters, which is influenced by previous research on program selection via majority voting such as CodeT~\cite{chen2022codet}, which demonstrate that, in the absence of additional information, selecting the solution with the highest estimated probability improves the likelihood of passing the tests.

\paragraph*{Program Repair Based \changes{Description} Repair}
If no semantic cluster matches all input-output examples, \projName applies program repair to fix faulty programs, using the given example inputs and outputs. Specifically, we employ the self-refine method~\cite{ding2024cycle}, which utilizes incorrect programs, together with the runtime feedback (e.g. test failures, expected output and actual output), and automatically generate a corrected program. \projName selects programs from the most probable cluster for repair, as this cluster represents the predominant interpretation by the LLM. \projName treats the faulty program as a rejected program, and the repaired program that is example consistent as a selected program. This process is provided in Algorithm~\ref{alg:overall_algorithm} in lines 19--20, \changes{and is illustrated in \Cref{fig:motivating-example3}, where a representative of the cluster with $\mathrm{EC}=0$ is repaired by removing the slant height computation so that it is consistent with the example, and then this change informs the necessary edit of the problem description.}

\subsection{Contrastive Specification Inference}

Once \projName has identified a selected program and one or more rejected programs, it applies our \emph{contrastive specification inference} prompt to disambiguate a given \changes{problem descriptions} via a minimal natural language change. Contrastive specification inference leverages the insight that an explanation is most informative when it highlights why outcome A occurs instead of outcome B rather than describing outcome A in isolation. By contrasting the behavior of the selected program with that of rejected programs, the repaired \changes{description} is prone to align with the selected program while intentionally diverge from the rejected ones. \projName realizes this objective by prompting an LLM with both the supporting evidence (selected program \& outputs) and the contradicting evidence (rejected programs \& outputs), and asking the model to (i) diagnose the sources of ambiguity and (ii) rewrite the specification so that only the intended behavior remains admissible. All prompts used by \projName are given in the supplementary materials.

\section{Evaluation}
\label{sec:evaluation}

\begin{table*}[htbp]
\centering
\caption{The effect of repairing \changes{problem descriptions} from HumanEval+, MBPP+, and \changes{LiveCodeBench} on code generation performance of four LLMs. Each approach is denoted as ``\{description repair method\} + \{code generation method\}''. ``Original'' refers to the \changes{descriptions} before repair. \projName consistently outperforms other approaches in code generation metrics and uncertainty reduction. ``Mod\%'' refers to the ratio of modified \changes{descriptions}.}
 \resizebox{\textwidth}{!}{
\setlength{\tabcolsep}{3pt}
\label{tab:rq1_overall}
\begin{tabular}{l l *{6}{c} *{6}{c} *{6}{c}}
\toprule
\textbf{Model} & \textbf{Approach} 
  & \multicolumn{6}{c}{\textbf{HumanEval+}} 
  & \multicolumn{6}{c}{\textbf{MBPP+}}
  & \multicolumn{6}{c}{\textbf{\changes{LiveCodeBench}}} \\
\cmidrule(lr){3-8} \cmidrule(lr){9-14} \cmidrule(lr){15-20}
 &  
  & Mod\% & Pass@1 & APR & NZP@1 & M@20 & SE 
  & Mod\% & Pass@1 & APR & NZP@1 & M@20 & SE
  & Mod\% & Pass@1 & APR & NZP@1 & M@20 & SE \\
\midrule
\multirow{5}{*}{DeepSeek} 
 & Original       
   & --       & 87.8\% & 94.1\% & \textbf{93.9\%} & 89.6\% & 0.1 
   & --       & 79.4\% & 88.9\% & \textbf{83.6\%} & 80.2\% & 0.1
   & --       & 37.5\% & 65.3\% & 44.6\% & 40.0\% & 0.5 \\
 & Original + $\mu$Fix        
   & --       & 88.9\% & 94.6\% & 93.3\% & 89.0\% & 0.1 
   & --       & 79.1\% & 89.4\% & 81.7\% & 79.7\% & 0.1
   & --       & 37.4\% & 65.7\% & 38.9\% & 37.5\% & \textbf{0.1} \\
 & Vanilla Repair  
   & 34.2\%       & 84.8\% & 92.2\% & 89.0\% & 86.6\% & 0.1 
   & 44.4\%       & 79.1\% & 88.5\% & 81.2\% & 79.4\% & 0.1
   & 33.5\%           & 35.2\% & 60.6\% & 42.3\% & 37.7\% & 0.5 \\
 & ClarifyGPT-auto  
   & 17.7\%       & 89.0\% & 95.2\% & 90.9\% & 89.8\% & 0.1 
   & 12.7\%       & 79.9\% & 89.9\% & 81.7\% & 80.0\% & 0.1
   & 61.7\%           & 38.1\% & 67.1\% & 47.0\% & 40.0\% & 0.4 \\
 & \textbf{\projName}    
   & 20.1\%       & \textbf{92.0\%} & \textbf{95.9\%} & 93.3\% & \textbf{92.7\%} & \textbf{0.0} 
   & 19.3\%       & \textbf{82.1\%} & \textbf{91.8\%} & 83.2\% & \textbf{82.2\%} & \textbf{0.0}
   & 66.3\%           & \textbf{40.8\%} & \textbf{69.9\%} & \textbf{47.0\%} & \textbf{41.9\%} & 0.3 \\
\midrule
\multirow{5}{*}{Qwen2.5} 
 & Original       
   & --       & 84.4\% & 92.7\% & 87.7\% & 85.2\% & 0.1 
   & --       & 79.2\% & 89.7\% & 81.5\% & 79.4\% & 0.1
   & --       & 29.4\% & 54.3\% & 40.0\% & 31.4\% & 0.5 \\ 
 & Original + $\mu$Fix        
   & --       & 83.9\% & 93.1\% & 85.0\% & 84.3\% & 0.0 
   & --       & 79.0\% & 89.1\% & 79.7\% & 78.9\% & 0.0
   & --       & 29.6\% & 54.9\% & 33.5\% & 30.1\% & \textbf{0.3} \\
 & Vanilla Repair  
   & 15.2\%       & 83.3\% & 92.3\% & 86.0\% & 84.1\% & 0.1 
   & 36.8\%       & 76.1\% & 86.8\% & 77.8\% & 76.2\% & 0.1
   & 1.8\%           & 29.3\% & 54.1\% & 39.4\% & 30.9\% & 0.5 \\
 & ClarifyGPT-auto    
   & 15.9\%       & 85.2\% & 93.9\% & 86.9\% & 85.6\% & 0.1 
   & 13.5\%       & 79.2\% & 89.5\% & 80.3\% & 79.1\% & 0.0
   & 69.1\%           & 28.9\% & 54.5\% & 37.3\% & 31.0\% & 0.5 \\
 & \textbf{\projName}      
   & 20.7\%       & \textbf{87.7\%} & \textbf{96.5\%} & \textbf{88.6\%} & \textbf{87.7\%} & \textbf{0.0} 
   & 16.9\%       & \textbf{82.3\%} & \textbf{91.9\%} & \textbf{83.5\%} & \textbf{82.7\%} & \textbf{0.0}
   & 77.1\%           & \textbf{33.1\%} & \textbf{60.1\%} & \textbf{40.6\%} & \textbf{34.9\%} & 0.4 \\
\midrule
\multirow{5}{*}{GPT-4o} 
 & Original       
   & --       & 82.0\% & 93.0\% & \textbf{91.5\%} & 83.5\% & 0.2 
   & --       & 78.2\% & 88.9\% & \textbf{84.9\%} & 78.8\% & 0.2
   & --       & 32.3\% & 54.2\% & 41.1\% & 33.1\% & 0.6 \\
 & Original + $\mu$Fix        
   & --       & 82.2\% & 93.0\% & 88.6\% & 83.9\% & 0.2 
   & --       & 78.0\% & 88.7\% & 81.1\% & 78.4\% & 0.2
   & --       & 30.5\% & 52.9\% & 35.0\% & 31.0\% & \textbf{0.3} \\
 & Vanilla Repair  
   & 62.8\%  & 80.1\% & 91.5\% & 84.1\% & 79.3\% & 0.2 
   & 82.5\%       & 76.0\% & 87.5\% & 82.0\% & 77.0\% & 0.2
   & 42.1\%           & 32.0\% & 54.7\% & 40.6\% & 32.6\% & 0.5 \\
 & ClarifyGPT-auto      
   & 36.0\%       & 82.3\% & 92.7\% & 87.8\% & 83.7\% & 0.1 
   & 30.2\%       & 77.6\% & 88.9\% & 80.7\% & 78.0\% & 0.1
   & 76.0\%           & 32.2\% & 55.0\% & 40.6\% & 33.7\% & 0.5 \\
 & \textbf{\projName}       
   & 39.6\%       & \textbf{87.6\%} & \textbf{96.4\%} & 90.2\% & \textbf{87.4\%} & \textbf{0.0} 
   & 31.8\%       & \textbf{80.6\%} & \textbf{91.2\%} & \textbf{82.7\%} & \textbf{80.8\%} & \textbf{0.0}
   & 82.3\%           & \textbf{35.6\%} & \textbf{58.5\%} & \textbf{41.7\%} & \textbf{37.9\%} & 0.3 \\
\midrule
\multirow{5}{*}{\changes{GPT-4o-mini}} 
 & Original       
   & --       & 81.6\% & 92.0\% & 92.7\% & 81.7\% & 0.2 
   & --       & 75.8\% & 86.6\% & 80.4\% & 76.7\% & 0.2
   & --       & 28.6\% & 53.6\% & 39.4\% & 31.4\% & 0.6 \\
 & Original + $\mu$Fix        
   & --       & 80.2\% & 90.4\% & 86.4\% & 81.7\% & 0.1 
   & --       & 76.0\% & 86.7\% & 79.0\% & 76.5\% & 0.1
   & --       & 28.2\% & 53.3\% & 32.0\% & 29.0\% & \textbf{0.4} \\
 & Vanilla Repair  
   & 11.8\%       & 81.5\% & 91.6\% & 91.5\% & 81.5\% & 0.2 
   & 31.3\%       & 75.5\% & 86.4\% & 79.6\% & 76.5\% & 0.2
   & 2.6\%       & 28.7\% & 53.4\% & 39.4\% & 31.4\% & 0.6 \\
 & ClarifyGPT-auto      
   & 30.5\%       & 80.1\% & 91.6\% & 86.0\% & 81.1\% & 0.1 
   & 31.0\%       & 75.1\% & 87.1\% & 77.3\% & 75.7\% & 0.1
   & 77.1\%       & 28.2\% & 53.3\% & 38.5\% & 30.3\% & 0.6 \\
 & \textbf{\projName}       
   & 34.1\%       & \textbf{87.6\%} & \textbf{95.3\%} & \textbf{93.5\%} & \textbf{88.0\%} & \textbf{0.1} 
   & 33.1\%       & \textbf{78.5\%} & \textbf{89.5\%} & \textbf{80.5\%} & \textbf{79.2\%} & \textbf{0.1}
   & 81.7\%       & \textbf{32.6\%} & \textbf{57.7\%} & \textbf{40.8\%} & \textbf{35.4\%} & 0.4 \\
\bottomrule
\end{tabular}
}
\vspace{-1mm}
\end{table*}

Our study addresses the following research questions:

\begin{itemize}
\item \textbf{RQ1}: To what extent do \projName-repaired \changes{descriptions} improve code generation compared to baseline methods?
\item \textbf{RQ2}: Do \changes{problem descriptions} repaired by \projName enable cross-model improvement in code generation?  
\item \textbf{RQ3}: How much does \projName change \changes{problem descriptions} in comparison with baseline methods?
\end{itemize}

\subsection{Experimental Setup}

\paragraph*{Models}
We selected four widely-used LLMs: GPT-4o, \changes{GPT-4o-mini}, DeepSeek-V3 and Qwen2.5-Coder-32b-Instruct. GPT-4o~\cite{hurst2024gpt} is OpenAI's advanced commercial model with high accuracy across diverse tasks, and \changes{GPT-4o-mini} is its smaller variant. DeepSeek-V3~\cite{liu2024deepseek} is an open-source, efficiency-focused model optimized for mathematics and coding. Qwen2.5-Coder-32B-Instruct~\cite{yang2024qwen2}, developed by Alibaba, is a smaller model that specializes in multilingual code generation, debugging, and software development with robust problem solving capabilities. For brevity, we refer to DeepSeek-V3 and Qwen2.5-Coder-32b-Instruct simply as ``DeepSeek'' and ``Qwen2.5''.

\paragraph*{Benchmarks}
We selected three widely-used code generation benchmarks: HumanEval+, MBPP+ \changes{and LiveCodeBench}.
HumanEval+ consists of 164 programming problems, each including a function signature, docstring, canonical solutions, and several test cases. These problems are designed to evaluate language comprehension, algorithms, and simple mathematics. 
MBPP+ includes 378 hand-verified Python programming problems, each with a task description, canonical solutions, and multiple test cases. These problems cover programming fundamentals and standard library functionalities.
HumanEval+ and MBPP+ are enhanced versions of the original HumanEval~\cite{chen2021evaluating} and MBPP~\cite{austin2021program} benchmarks augmented by the EvalPlus framework~\cite{liu2023your}. Humaneval+ includes 2.81 examples per problem on average and 161 \changes{problem descriptions} embed at least one example. MBPP+ includes one example in each problem.
\changes{LiveCodeBench~\cite{jain2025livecodebench} is a dynamic, contamination-aware benchmark that continuously harvests recent competitive-programming problems from LeetCode, AtCoder, and Codeforces. To minimize the risk of data leakage, we select livecodebench\_v6, which contains 175 problems released between January 2025 and May 2025.}
The hidden tests used to assess generated programs are different from the examples in the \changes{problem descriptions}.


Following prior work~\cite{mu2023clarifygpt}, we fix the model temperature at 0 for all tasks to ensure deterministic outputs, except for program sampling, where the temperature is set to the default value specified by each LLM to encourage diversity. \changes{Since our approach, just like most works on uncertainty estimation, relies on diversity in the output, it requires non-zero sampling temperature.} To avoid suggesting that disambiguation was always successful, we use the term ``modified'' rather than ``repaired'' in certain descriptions. To account for generation variability, we repeated all experiments three times and reported the average performance.

\changes{To estimate entropy, we sampled $N$ programs, thus it is necessary to choose $N$ that provides a reliable estimate. We conducted an experiment on a subset of problems with various $N$ $\in$ {5, 10, 15, ..., 40}, which showed that entropy increases rapidly from 0.096 (N=5) to 0.1123 (N=20), after which additional samples result only in small fluctuations. Therefore, we adopt N=20 as our default configuration, ensuring both statistical reliability and computational efficiency.}

\subsection{Evaluated Approaches}

\changes{To the best of our knowledge, \projName is the first approach designed specifically for repairing problem descriptions for LLM-based code generation in a fully-automated fashion.} Thus, for our evaluation, we adapted the following approaches to \changes{problem description} repair to serve as baselines:

\textbf{Vanilla Repair} is a baseline \changes{problem description} repair approach: an LLM first classifies each \changes{description} as ambiguous or unambiguous. If a \changes{description} is judged as ambiguous, the model is prompted to resolve the ambiguity by adopting the most likely interpretation.


\textbf{ClarifyGPT-auto} adapts ClarifyGPT~\cite{mu2023clarifygpt} for fully automated \changes{problem description} repair. Whereas the original approach simulates human clarifications using hidden test cases, we replace this mechanism with examples drawn directly from the \changes{problem descriptions} to eliminate data‐leakage risks. The resulting ``repaired'' \changes{description} is formed by concatenating the original specification with the model’s clarifications. Semantic clustering again uses $N=20$ samples.



\textbf{$\mu$Fix}~\cite{tian2025fixing} analyzes program failures on input–output examples and generates reasoning chains to iteratively refine its interpretation of \changes{problem descriptions}. Although $\mu$Fix does not produce explicit \changes{description} edits, its focus on improving \changes{problem} understanding is relevant to \projName.

\begin{table*}[t]
  \centering
  \small
  \caption{For each baseline (Vanilla Repair, ClarifyGPT-auto and $\mu$Fix), this table compares performance of \projName's repairs with the baseline's performance on the subset of \changes{problem descriptions} that both \projName and the baseline modified. ``\#'' denotes the number of such \changes{descriptions}. Since $\mu$Fix does not modify \changes{descriptions}, the corresponding subsets includes all \changes{descriptions} \projName modified. \projName-repaired \changes{descriptions} outperform baselines on most subsets.}
  \label{tab:intersection_evaluation}
   \resizebox{\textwidth}{!}{
  \setlength{\tabcolsep}{5pt}
  \begin{tabular}{@{}l l
                  r
                  r
                  r
                  r
                  r
                  r
                  r
                  r
                  r
                  r
                  r
                  r
                  @{}}
    \toprule
    \multirow{3}{*}{\textbf{Model}} 
      & \multirow{3}{*}{\textbf{Baseline}}
      & \multicolumn{4}{c}{\textbf{HumanEval+}} 
      & \multicolumn{4}{c}{\textbf{MBPP+}}
      & \multicolumn{4}{c}{\textbf{\changes{LiveCodeBench}}} \\
    \cmidrule(lr){3-6}\cmidrule(lr){7-10}\cmidrule(lr){11-14}
      & 
      & \multirow[c]{2}{*}{\textbf{\#}} 
      & \multicolumn{3}{c}{\textbf{Pass@1}}
      & \multirow[c]{2}{*}{\textbf{\#}} 
      & \multicolumn{3}{c}{\textbf{Pass@1}}
      & \multirow[c]{2}{*}{\textbf{\#}} 
      & \multicolumn{3}{c}{\textbf{Pass@1}} \\
    \cmidrule(lr){4-6}\cmidrule(lr){8-10}\cmidrule(lr){12-14}
      & 
      & 
      & \textbf{Original} 
      & \textbf{Baseline} 
      & \textbf{\projName}
      & 
      & \textbf{Original} 
      & \textbf{Baseline} 
      & \textbf{\projName}
      &
      & \textbf{Original}
      & \textbf{Baseline}
      & \textbf{\projName} \\
    \midrule
    \multirow{3}{*}{Deepseek}
      & Vanilla Repair & 17 & 34.12\% & 29.41\% & \textbf{51.76\%} &  48 & 38.75\% & 39.58\% & \textbf{46.74\%} & 40 & 9.06\% & 7.00\% & \textbf{12.93\%} \\
      & ClarifyGPT-auto     & 29 & 48.62\% & 55.63\% & \textbf{71.15\%} &  64 & 42.66\% & 45.83\% & \textbf{52.08\%} & 108 & 15.53\% & 16.81\% & \textbf{20.28\%} \\
      & Original + $\mu$Fix          & 33 & 45.76\% & 54.34\% & \textbf{66.57\%} &  73 & 37.67\% & 43.47\% & \textbf{52.05\%} & 116 & 16.08\% & 16.50\% & \textbf{21.09\%} \\
    \midrule
    \multirow{3}{*}{Qwen2.5}
      & Vanilla Repair &  8 & \textbf{36.25\%} & 35.00\% & 29.17\% &  29 & 53.21\% & 42.50\% & \textbf{62.02\%} & 3 & 6.67\% & 0.00\% & \textbf{22.22\%} \\
      & ClarifyGPT-auto      & 26 & 49.66\% & 54.23\% & \textbf{64.36\%}&  51 & 50.20\% & 50.63\% & \textbf{57.89\%} & 121 & 13.01\% & 12.36\% & \textbf{18.30\%} \\
      & Original + $\mu$Fix          & 34 & 47.97\% & 63.91\% & \textbf{64.71\%} &  64 & 40.51\% & 51.30\% & \textbf{60.17\%} & 135 & 12.41\% & 13.57\% & \textbf{17.23\%} \\
    \midrule
    \multirow{3}{*}{GPT-4o}
      & Vanilla Repair& 50 & 50.82\% & 53.76\% & \textbf{70.20\%} & 104 & 54.93\% & 53.45\% & \textbf{61.15\%} & 60 & 14.41\% & 12.96\% & \textbf{21.07\%} \\
      & ClarifyGPT-auto      & 59 & 61.21\% & 62.08\% & \textbf{73.61\%} & 114 & 55.46\% & 53.37\% & \textbf{61.05\%} & 133 & 18.19\% & 18.94\% & \textbf{22.98\%} \\
      & Original + $\mu$Fix          & 65 & 58.17\% & 63.81\% & \textbf{72.30\% }& 120 & 52.69\% & 52.53\% & \textbf{60.03\%} & 144 & 21.33\% & 19.74\% & \textbf{25.43\%} \\
    \midrule
    \multirow{3}{*}{\changes{GPT-4o-mini}}
      & Vanilla Repair & 7 & 45.71\% & 44.29\% & \textbf{61.43\%} & 25 & 32.80\% & 30.00\% & \textbf{46.40\%} & 2 & 35.00\% & 50.00\% & \textbf{50.00\%} \\
      & ClarifyGPT-auto & 50 & 57.36\% & 52.46\% & \textbf{74.56\%} & 117 & 47.01\% & 44.44\% & \textbf{53.68\%} & 135 & 16.64\% & 16.12\% & \textbf{21.82\%} \\
      & Original + $\mu$Fix & 56 & 56.75\% & 58.33\% & \textbf{74.54\%} & 125 & 44.00\% & 47.70\% & \textbf{52.16\%} & 143 & 16.29\% & 15.76\% & \textbf{21.10\%} \\
    \bottomrule
  \end{tabular}
  }
  \vspace{-4mm}
\end{table*}

\subsection{Evaluation Metrics}
We apply five measures to comprehensively evaluate the correctness and interpretive clarity of repaired \changes{descriptions}.

\textbf{Pass@1} is the probability that a generated program passes all tests. We estimate Pass@1 by sampling ten independent programs per problem according to the best practices~\cite{chen2021evaluating}.

\textbf{AvgPassRate} (APR) measures the expected ratio of tests passed across generated programs.
APR provides a more granular view of how disambiguating \changes{problem descriptions} improves partial correctness.

\textbf{\%Pass@1$>$0} (NZP@1) denotes the proportion of problems for which Pass@1 exceeds zero. 
An increase in NZP@1 when comparing original versus repaired \changes{problem descriptions} indicates that repair enables the model to produce at least one fully correct program for cases previously unsolvable.

\textbf{Majority@20} (M@20) measures the accuracy of the most frequent correct program within 20 generated samples (i.e. the majority vote)~\cite{chen2022codet,li2022competition}. This metric evaluates the model’s consistency in converging on a consensus solution.

\textbf{Semantic Entropy} (SE)~\cite{kuhn2023semantic} quantifies the distribution of generated programs across distinct semantic clusters. Higher semantic entropy suggests greater ambiguity~\cite{mu2023clarifygpt}.

\begin{table}[t]
\centering
\footnotesize
\setlength{\tabcolsep}{1.2pt}
\renewcommand{\arraystretch}{1.1}
\caption{Pass@1 (\%) on HumanEval+, MBPP+, and LiveCodeBench before (“Orig.”) and after (“Repair”) ambiguity repair. Rows are the description-repair models that perform the repair; columns are the models used for evaluation. Each cell shows Orig./Repair. Underlined cells indicate the same model is used for both repair and evaluation. “\#” is the number of modified \changes{descriptions}.
Underlined cells indicate the same model performed both repair and evaluation.}
\label{tab:generalization}

\textbf{HumanEval+}\vspace{2pt}

\begin{tabularx}{\columnwidth}{@{}l c *{4}{>{\centering\arraybackslash}X}@{}}
\toprule
\textbf{Repair Model} & \textbf{\#} &
\textbf{DeepSeek} & \textbf{Qwen2.5} & \textbf{GPT-4o} & \textbf{\changes{GPT-4o-mini}} \\
\midrule
DeepSeek & 33 &
\underline{45.76/\textcolor{darkgreen}{66.57}} &
51.55/\textcolor{darkgreen}{61.10} &
44.88/\textcolor{darkgreen}{56.18} &
45.69/\textcolor{darkgreen}{60.71} \\
Qwen2.5 & 34 &
60.59/\textcolor{darkgreen}{65.00} &
\underline{43.86/\textcolor{darkgreen}{59.12}} &
49.15/\textcolor{darkgreen}{58.05} &
46.11/\textcolor{darkgreen}{52.65} \\
GPT-4o & 65 &
72.62/\textcolor{red}{71.13} &
67.88/\textcolor{darkgreen}{74.04} &
\underline{58.17/\textcolor{darkgreen}{72.30}} &
61.20/\textcolor{darkgreen}{69.44} \\
\changes{GPT-4o-mini} & 56 &
71.79/\textcolor{darkgreen}{72.26} &
66.69/\textcolor{darkgreen}{73.14} &
65.91/\textcolor{darkgreen}{73.34} &
\underline{56.75/\textcolor{darkgreen}{74.54}} \\
\bottomrule
\end{tabularx}

\vspace{6pt}
\textbf{MBPP+}\vspace{2pt}

\begin{tabularx}{\columnwidth}{@{}l c *{4}{>{\centering\arraybackslash}X}@{}}
\toprule
\textbf{Repair Model} & \textbf{\#} &
\textbf{DeepSeek} & \textbf{Qwen2.5} & \textbf{GPT-4o} & \textbf{\changes{GPT-4o-mini}} \\
\midrule
DeepSeek & 73 &
\underline{37.67/\textcolor{darkgreen}{52.05}} &
49.95/\textcolor{red}{46.68} &
46.30/\textcolor{darkgreen}{49.96} &
42.05/\textcolor{darkgreen}{46.58} \\
Qwen2.5 & 64 &
47.03/\textcolor{darkgreen}{54.74} &
\underline{40.00/\textcolor{darkgreen}{58.12}} &
46.45/\textcolor{darkgreen}{54.72} &
44.69/\textcolor{darkgreen}{46.30} \\
GPT-4o & 120 &
55.75/\textcolor{darkgreen}{60.03} &
58.41/\textcolor{darkgreen}{61.51} &
\underline{52.69/\textcolor{darkgreen}{60.03}} &
48.58/\textcolor{darkgreen}{55.36} \\
\changes{GPT-4o-mini} & 125 &
55.36/\textcolor{red}{51.03} &
57.11/\textcolor{red}{51.42} &
52.10/\textcolor{red}{48.70} &
\underline{44.00/\textcolor{darkgreen}{52.16}} \\
\bottomrule
\end{tabularx}

\vspace{6pt}
\textbf{LiveCodeBench}\vspace{2pt}

\begin{tabularx}{\columnwidth}{@{}l c *{4}{>{\centering\arraybackslash}X}@{}}
\toprule
\textbf{Repair Model} & \textbf{\#} &
\textbf{DeepSeek} & \textbf{Qwen2.5} & \textbf{GPT-4o} & \textbf{\changes{GPT-4o-mini}} \\
\midrule
DeepSeek     & 116 & 16.08/\textcolor{darkgreen}{32.09} & 9.82/\textcolor{darkgreen}{12.91} & 12.60/\textcolor{darkgreen}{15.13} & 9.14/\textcolor{darkgreen}{11.61} \\
Qwen2.5      & 135 & 22.35/\textcolor{darkgreen}{23.00} & 12.26/\textcolor{darkgreen}{17.06} & 17.80/\textcolor{darkgreen}{18.29} & 12.66/\textcolor{darkgreen}{15.28} \\
GPT-4o       & 144 & 26.91/\textcolor{darkgreen}{27.39} & 18.73/\textcolor{darkgreen}{23.52} & 21.33/\textcolor{darkgreen}{25.43} & 17.68/\textcolor{darkgreen}{20.62} \\
\changes{GPT-4o-mini}  & 143 & 27.64/\textcolor{darkgreen}{28.17} & 17.28/\textcolor{darkgreen}{20.86} & 22.12/\textcolor{red}{20.96} & \underline{16.29/\textcolor{darkgreen}{21.10}} \\
\bottomrule
\end{tabularx}
\vspace{-3mm}
\end{table}

\subsection{\textbf{RQ1:} Repair Effectiveness of \projName}

To evaluate the effectiveness of generated repairs, we measured how \changes{problem descriptions} modified by an LLM affect the same LLM's performance. Table~\ref{tab:rq1_overall} presents the ratio of modified \changes{ descriptions} and performance metrics on all problems in benchmarks. 
For every model–benchmark pairing, \projName achieves the highest Pass@1 score among all approaches. For instance, on the HumanEval+ benchmark with DeepSeek, \projName achieves a Pass@1 of 92.0\% surpassing the performance of the original \changes{descriptions} (87.8\%), and the results of ClarifyGPT-auto (89.0\%). Similar improvements are observed for Qwen2.5, GPT-4o and GPT-4o-mini. In addition, \projName achieves the highest average pass rate, with DeepSeek/HumanEval+ yielding an AvgPassRate of 95.6\%. This suggests \projName repairs guide models toward outputs passing a higher ratio of tests. Notably, \projName also substantially reduces semantic entropy, showing that repaired \changes{descriptions} yield less semantically diverse code. Finally, \projName is the only approach that consistently improved Majority@20, showing that \projName not only increases the probability of generating a correct program but also enhances the consistency with which a consensus solution emerges under repeated sampling.

An alternative to improving the quality of \changes{problem descriptions} is enhancing LLM's code generation capabilities to alleviate shortcomings of the \changes{problem descriptions}. Thus, we investigate how the performance improvement due to \changes{descriptions} repaired by \projName in the context of zero-shot code generation compares to the performance of original \changes{descriptions} with a SOTA code generation method, $\mu$Fix, which employs advanced reasoning techniques. Table~\ref{tab:rq1_overall} present the comparative results.  Since $\mu$Fix reasons about all of the benchmark problems, it does not modify a subset of them, hence no ``Modified\%'' is shown for this method in Table~\ref{tab:rq1_overall}. \projName outperformed $\mu$Fix in most settings. Moreover, $\mu$Fix showed only minimal enhancements in code generation measures compared to the zero-shot approach. We attribute this to the fact that benchmarks like HumanEval and MBPP are reaching their saturation~\cite{xia2024top}, where SOTA models already exhibit high success rate, making additional improvements through prompt tuning challenging. \projName's repaired \changes{descriptions} demonstrated substantial performance gains, showing that \changes{problem description} repair is a promising pathway for further pushing the boundaries of code generation.

We also investigated Pass@1 only on the modified \changes{problem descriptions} shown in Table~\ref{tab:generalization}. On average, 43.58\% of \changes{descriptions} are modified by \projName. The underlined values illustrate the Pass@1 results when the same model performs both repair and evaluation. All models exhibit improvements, with an average Pass@1 increase of 30.9\%.

Since each method uses a different heuristic to detect ambiguity, they operate on partly disjoint sets of \changes{problem descriptions}. Meanwhile, Monperrus~\cite{monperrus2014critical} argued that an objective comparison of repair methods needs to consider the classes of defects the methods address. In order to eliminate the influence brought by different ambiguity detection strategies,  we performed pairwise evaluations of \projName against each baseline, restricted to the intersection of \changes{problem descriptions} they both modify. Table~\ref{tab:intersection_evaluation} shows the Pass@1 on these subsets of \changes{problem descriptions}. \projName yields higher Pass@1 than both the original \changes{description} and the baseline repair methods on all subsets except for the \changes{descriptions} jointly repaired with Vanilla Repair on HumanEval+ with Qwen2.5. We attribute this to the fact that the set is very small (8 cases) and the performance drop might be the result of statistical noise.


\changes{\projName's average repair time is 36.7 seconds, and the average token cost is 6770.7 tokens per description.}

\begin{rqBox}{1}{Across benchmarks, \projName yields a 30.9\% average improvement in Pass@1 on the modified subset, corresponding to a 4.09\% improvement over complete benchmarks. Baseline methods commonly decreased performance or produced negligible gains ($<$0.5\%).}
\end{rqBox}

\subsection{\textbf{RQ2}: Cross-Model Generalization}

To show that \projName addresses inherent problems of \changes{problem descriptions} rather than merely model-specific misunderstandings, we evaluate how \changes{descriptions} repaired using one model affect another model's code generation performance. \Cref{tab:generalization} shows the results of our experiments. Although in few instances, which are highlighted in red, the other model's performance dropped after repair, specifically MBPP+ repaired by GPT-4o-mini, on average using a different model to repair \changes{problem descriptions} increases performance by 10.48\%.

\begin{rqBox}{2}{\changes{Problem descriptions} repaired by \projName using one model improve another model’s Pass@1 by 10.48\% on average, showing that its repairs generalize across models.}
\end{rqBox}

\subsection{\textbf{RQ3}: Comparing Lengths of Repaired \changes{Descriptions}}
To further illustrate the limitations of baseline methods in repairing ambiguous \changes{problem descriptions},  we measured the relative increase in specification length induced by each method.
Table~\ref{tab:requirement_length} reports the average percentage growth of the repaired \changes{problem descriptions}. ClarifyGPT-auto's clarification and $\mu$Fix's reasoning chains both significantly increased the \changes{descriptions} length. When using GPT-4o on MBPP+, the post-repair \changes{descriptions} generated by ClarifyGPT-auto and $\mu$Fix respectively increased by 576.61\% and 425.56\% compared to the original ones. Although Vanilla Repair yields the smallest length increase, it also achieves the weakest repair effectiveness. In contrast, \projName achieves the best repair performance while also ensuring the \changes{description} simplicity after the repair.
ClarifyGPT-auto generates irrelevant clarifying questions so the repaired \changes{problem descriptions} include redundant clarifications. $\mu$Fix's format-specific prompts embed full reasoning chains in the repaired \changes{descriptions}. Introducing too many explanations will not only reduce the model's understanding of the original problem, but also may lead to the model overfitting to clarifications or examples.

\begin{rqBox}{3}{\projName’s repairs result in only modest increases in \changes{description} length compared with ClarifyGPT-auto's clarifications and $\mu$Fix's reasoning chains, while still achieving superior repair effectiveness.}

\end{rqBox}


\begin{table}[t]
  \centering
  \setlength{\tabcolsep}{4pt}
  \renewcommand{\arraystretch}{1.12}
  \caption{Relative increment in \changes{description} length between before‐repair and after‐repair for HumanEval+, MBPP+ and \changes{LiveCodeBench}. LCB refers to LiveCodeBench.}
  \label{tab:requirement_length}
  \resizebox{\columnwidth}{!}{%
  \begin{tabular}{@{} l l
      S[table-format=+3.1]
      S[table-format=+3.1]
      S[table-format=+3.1] @{}}
    \toprule
    \multicolumn{1}{l}{\textbf{Model}} &
    \multicolumn{1}{l}{\textbf{Method}} &
    \multicolumn{1}{c}{\textbf{HumanEval+}} &
    \multicolumn{1}{c}{\textbf{MBPP+}} &
    \multicolumn{1}{c}{\textbf{\changes{LCB}}} \\
    \midrule

    \multirow{4}{*}{DeepSeek}
      & Vanilla Repair  &  -3.9\% &  +57.4\% &  -46.4\% \\
      & ClarifyGPT-auto      & +158.8\% & +360.7\% &  +64.4\% \\
      & $\mu$fix        & +255.0\% & +354.0\% & +339.9\% \\
      & \textbf{\projName}        &  +12.8\% & +120.2\% &   +0.7\% \\
    \midrule

    \multirow{4}{*}{Qwen2.5}
      & Vanilla Repair  &  -44.9\% &  -31.5\% &  -62.5\% \\
      & ClarifyGPT-auto      &  +82.5\% & +184.8\% &  +32.9\% \\
      & $\mu$fix        & +405.3\% & +795.2\% & +339.3\% \\
      & \textbf{\projName}         &   +9.0\% &  +58.4\% &   +8.0\% \\
    \midrule

    \multirow{4}{*}{GPT-4o}
      & Vanilla Repair  &   -3.9\% &   +2.2\% &  -37.7\% \\
      & ClarifyGPT-auto      & +262.2\% & +576.6\% &  +94.3\% \\
      & $\mu$fix        & +298.6\% & +426.6\% & +375.8\% \\
      & \textbf{\projName}         &  +65.8\% & +237.4\% &  +43.9\% \\
    \midrule

    \multirow{4}{*}{\changes{GPT-4o-mini}}
      & Vanilla Repair  &  +13.3\% &  +10.2\% &  -58.0\% \\
      & ClarifyGPT-auto      & +114.0\% & +281.1\% &  +23.1\% \\
      & $\mu$fix        & +263.5\% & +317.4\% & +243.4\% \\
      & \textbf{\projName}         &  +22.0\% & +113.7\% &  +10.9\% \\
    \bottomrule
  \end{tabular}
  }
  \vspace{-4mm}
\end{table}

\subsection{Ablation Study}
\changes{\projName uses two measures to guide repair: semantic entropy (SE) and example consistency (EC). To investigate their contributions, we compared \projName with $\projName_{woSE}$ (removing SE) and $\projName_{woEC}$ (removing EC). Because \projName and its two variants modify different descriptions under their own criteria, we report the Pass@1 over the entire datasets. Table~\ref{tab:ablation_study} reports the Pass@1 across 3 datasets and 4 models in terms of Pass@1. First, \projName consistently outperforms both $\projName_{woSE}$ and $\projName_{woEC}$, indicating both two components is beneficial. On average, \projName improves 2.2\% and 0.92\% higher Pass@1 than $\projName_{woSE}$ and $\projName_{woEC}$, respectively.}


\changes{Finally, \projName involves one hyperparameter $k$ (the number of repair iteration), for which we investigated different settings ($k$ $\in$ 1, 2,..., 10) on LiveCodeBench and Qwen2.5. Pass@1 increases from 31.17\% (k=1) to 36.51\% (k=3), after which further iterations produce only minor fluctuations.}

\subsection{Analysis and Mitigation of Repair Errors}

\changes{We conducted an analysis of cases where \projName generated incorrect modifications. Table~\ref{tab:mitigation} reports the incorrect modification ratio (IMR), i.e. the fraction of modifications that reduce Pass@1, and their corresponding Pass@1. On average, IMR is 3.23\%, and Pass@1 on incorrectly modifications is 20.93\%. Manual inspection shows most errors arise due to majority voting: when a correct and an incorrect clusters have similar probabilities, majority vote often selects an incorrect one, leading to faulty modifications. As a mitigation strategy, we apply modified z-score for majority voting: a cluster is selected only if its probability substantially exceeds all others~\cite{iglewicz1993volume}. Otherwise, \projName defers to user input. In comparison to ClarifyGPT that required the user to answer clarifying questions for each description, we only ask user to choose between two programs in 10.8\% of modifications. We simulated the user using hidden tests by selecting the cluster that passes more hidden tests. This resulted in IMR's drop to 1.58\% and Pass@1's rise to 47.93\%.}

\changes{Apart from the limitation of majority-voting, \projName operates under the assumption that the input–output examples are correct, which is a standard assumotion in test-based automated program repair research. In HumanEval+ and MBPP+, we identified 5 out of 542 instances where the examples conflicted with the reference solutions ($<$ 1\%), causing \projName to produce incorrect clarifications. This is small relative to the description modified by SpecFix (24\%).}

\begin{table}[t]
  \centering
  \caption{Contribution of semantic entropy (SE) and example consistency (EC) to Pass@1 (\%) of \projName's repairs.}
  \label{tab:ablation_study}
  \footnotesize
  \setlength{\tabcolsep}{4pt}
  \renewcommand{\arraystretch}{1.15}
  \begin{tabularx}{\columnwidth}{@{} l l *{3}{>{\centering\arraybackslash}X} @{}}
    \toprule
    \textbf{Dataset} & \textbf{Model} & \textbf{$\projName_{woSE}$} & \textbf{$\projName_{woEC}$} & \textbf{\projName} \\
    \midrule
    \multirow{4}{*}{\textbf{HumanEval+}}
      & DeepSeek-V3   & 90.55 & 89.95 & \textbf{91.99} \\
      & Qwen2.5       & 87.23 & 86.77 & \textbf{87.74} \\
      & GPT-4o        & 86.57 & 84.07 & \textbf{87.62} \\
      & \changes{GPT-4o-mini} & 86.22 & 84.33 & \textbf{87.65} \\
    \midrule
    \multirow{4}{*}{\textbf{MBPP+}}
      & DeepSeek-V3   & 81.53 & 79.87 & \textbf{82.14} \\
      & Qwen2.5       & 81.06 & 78.02 & \textbf{82.31} \\
      & GPT-4o        & 80.22 & 78.28 & \textbf{80.56} \\
      & \changes{GPT-4o-mini} & 77.41 & 76.98 & \textbf{78.54} \\
    \midrule
    \multirow{4}{*}{\textbf{\changes{LCB}}}
      & DeepSeek-V3   & 39.80 & 38.77 & \textbf{40.84} \\
      & Qwen2.5       & 32.88 & 32.14 & \textbf{33.12} \\
      & GPT-4o        & 34.84 & 34.23 & \textbf{35.62} \\
      & \changes{GPT-4o-mini} & 31.34 & 30.63 & \textbf{32.55} \\
    \bottomrule
  \end{tabularx}
  \vspace{-4mm}
\end{table}

\begin{table}[t]
\centering
\caption{Incorrect-Modification Rate (IMR) and corresponding Pass@1 (\%) across models and benchmarks. Entries are reported as \emph{before mitigation}\(\rightarrow\)\emph{after mitigation}; arrows indicate the direction of change after mitigation (↓ = lower IMR, ↑ = higher Pass@1). IMR is the fraction of modified descriptions that are incorrect.}
\label{tab:mitigation}
\resizebox{\columnwidth}{!}{%
\begin{tabular}{l l c c}
\toprule
\textbf{Model} & \textbf{Benchmark} & \textbf{IMR} & \textbf{Pass@1} \\
\midrule
\multirow{3}{*}{DeepSeek-V3}
 & HumanEval & \(3.0\%\to2.4\%\ \downarrow\)  & \(0.0\%\to10.0\%\ \uparrow\) \\
 & MBPP   & \(3.3\%\to2.9\%\ \downarrow\)  & \(3.8\%\to24.8\%\ \uparrow\) \\
 & \changes{LCB}   & \(2.5\%\to1.3\%\ \downarrow\)  & \(31.1\%\to50.3\%\ \uparrow\) \\
\midrule
\multirow{3}{*}{GPT-4o}
 & HumanEval & \(3.4\%\to0.9\%\ \downarrow\)   & \(17.3\%\to64.5\%\ \uparrow\) \\
 & MBPP     & \(7.2\%\to4.7\%\ \downarrow\)   & \(9.0\%\to31.4\%\ \uparrow\) \\
 & \changes{LCB}       & \(2.3\%\to1.4\%\ \downarrow\)   & \(33.3\%\to51.6\%\ \uparrow\) \\
\midrule
\multirow{3}{*}{\changes{GPT-4o-mini}}
 & HumanEval & \(2.2\%\to0.8\%\ \downarrow\)   & \(33.3\%\to53.6\%\ \uparrow\) \\
 & MBPP     & \(2.6\%\to1.6\%\ \downarrow\)   & \(20.1\%\to36.6\%\ \uparrow\) \\
 & \changes{LCB}       & \(3.0\%\to1.0\%\ \downarrow\)   & \(40.0\%\to57.4\%\ \uparrow\) \\
\midrule
\multirow{3}{*}{Qwen2.5}
 & HumanEval & \(3.9\%\to0.7\%\ \downarrow\)   & \(10.4\%\to60.0\%\ \uparrow\) \\
 & MBPP      & \(2.3\%\to0.0\%\ \downarrow\)   & \(25.7\%\to91.7\%\ \uparrow\) \\
 & \changes{LCB}      & \(3.0\%\to1.2\%\ \downarrow\)   & \(27.1\%\to43.3\%\ \uparrow\) \\
\bottomrule
\end{tabular}}
\end{table}

\section{Threats to Validity}

\changes{\projName assumes that code generated from problem descriptions can be independently executed, and that some candidate programs are correct or near-correct. However, this assumption may not hold, e.g. when this code is a part of a larger system or when the task’s complexity prevents the model from producing meaningful solutions. In such cases, ambiguity could instead be detected through behavioral proxies, such as formal models, which requires further investigation.}

Construct validity may be threatened by our operationalization of ambiguity and its repairs in terms of various metrics, such as Pass@1, semantic entropy, example consistency, etc. We adopted this approach because our goal was to investigate a fully-automated method of ambiguity resolution. However, resolving some ambiguities might require human input. In future research, we will investigate how to efficiently involve humans in the SpecFix workflow.

External validity is limited by our focus on HumanEval+, MBPP+ and LiveCodeBench, and relying on the input-output examples embedded in their \changes{problem descriptions}. These benchmarks are well-established and representative, and using input-output examples is a common practice, e.g. in programming-by-example~\cite{le2014flashextract}, in Stack Overflow posts~\cite{uddin2020mining} and GitHub issues~\cite{hirsch2021identifying}. However, studying ambiguities in real-world applications such as in interactions with AI coding assistants remains an important future research direction.

\section{Related Work}

\paragraph*{LLM Reasoning About \changes{Problem Descriptions}}
$\mu$FIX~\cite{tian2025fixing} focuses on enhancing \changes{problem description} understanding by integrating thought-eliciting prompting with feedback-based code generation. Fan et al.~\cite{fan2024self} applies LLM reasoning to elaborate the meaning of low-frequency keywords in \changes{problem descriptions}. In contrast, SpecFix aims to disambiguate the \changes{problem descriptions} themselves so that the models interpret them correctly. Our experiments show that SpecFix is more effective than $\mu$FIX in interpreting non-trivial input-output examples.
Li et al.~\cite{li2024generating} proposed maintaining equivalent representations, such as natural language comments and pseudocode, that preserve semantics, using a reflection mechanism with two LLMs. This approach may enhance SpecFix' reflection prompt used for contrastive inference.
CodeMind~\cite{liu2024codemind} introduces a ``specification reasoning'' task that evaluates model's ability to reason about combinations of NL specifications and test execution. Our study reveals that SOTA LLMs often perform poorly on this task, often generating program contradicting to explicitly specified examples.


\paragraph*{Detecting NL Ambiguity with LLMs} Ambiguity in \changes{problem descriptions} undermines downstream performance and has long attracted attention. \changes{A problem description for LLM-based code generation is a special case of software requirements}. Pre–ChatGPT work includes NLP‐based detection of inconsistency and vagueness~\cite{ezzini2022automated,luitel2023using,ferrari2019nlp,ferrari2018detecting} in software requirements, while recent efforts address semantic uncertainty in the ChatGPT era~\cite{fantechi2023inconsistency}. Vijayvargiya et al.\cite{vijayvargiya2025interactive} quantify the impact of ambiguous LLM inputs, showing up to a 20\% degradation in LLM performance. ClarifyGPT\cite{mu2023clarifygpt} measures uncertainty in generated code to trigger clarifying questions; although promising, it often produces irrelevant queries and fails to incorporate embedded examples. \projName overcomes these shortcomings by repairing the induced program distribution first and then inferring concise \changes{description} edits via contrastive specification inference.

\paragraph*{Prompt Optimization}
Prompt optimization is closely related to \changes{problem descriptions} repair.
Ma et al.~\cite{ma2024large} demonstrate that reflection‐based prompt tuning frequently fails to identify the root causes of prompt errors. SpecFix addresses this limitation by analyzing and repairing program distributions.

\paragraph*{Confidence and Uncertainty}
Hou et al.~\cite{hou2023decomposing} show connection between task ambiguity with aleatoric uncertainty, i.e. uncertainty resulting from inherent randomness in the data-generating process. The uncertainty can be estimated via semantic entropy~\cite{kuhn2023semantic}, which in code generation can measured by partitioning generated program into equivalence classes via differential testing~\cite{sharma2025assessing,fan2024oracle,mu2023clarifygpt}, a method adapted by SpecFix.

\paragraph*{Specification Inference with LLMs}
SpecRover~\cite{ruan2024specrover} infers program specification expressed in natural language to generate bugfixes. Multiple approaches apply LLMs to infer formal specifications from text~\cite{ma2024specgen,endres2024can}. The key novelty of SpecFix in comparison with these techniques is in its contrastive specification inference that infers specification that differentiates two programs.

\paragraph*{Automated Program Repair}
Program repair~\cite{le2019automated} aims to modify a given program to meet given correctness criteria. It is embodied in heuristics-based~\cite{le2011genprog}, semantics-based~\cite{mechtaev2016angelix}, and more recently in ML/LLM-based techniques~\cite{xia2023keep,bouzenia2024repairagent,parasaram2024fact}. Our innovating is in repairing \changes{problem descriptions} instead of programs, however, inspired by Fan et al.~\cite{fan2023automated}, we use program repair as a part of our algorithm.

\section{Conclusion}

This paper introduces the problem of automated repair of ambiguous \changes{problem descriptions for LLM-based code generation}. It shows that \changes{descriptions} can be repaired in a fully-automated fashion by aligning natural language with input-output examples and reducing uncertainty of code generation. \projName accomplishes this by analyzing and repairing the distribution of programs the \changes{description} induces, and then mapping the change back to the \changes{description}. Our experimental evaluation shows that repaired \changes{problem descriptions} significantly improve code generation performance of SOTA LLMs.



\bibliographystyle{IEEEtran}
\bibliography{refs}

\begin{thebibliography}{10}
\providecommand{\url}[1]{#1}
\csname url@samestyle\endcsname
\providecommand{\newblock}{\relax}
\providecommand{\bibinfo}[2]{#2}
\providecommand{\BIBentrySTDinterwordspacing}{\spaceskip=0pt\relax}
\providecommand{\BIBentryALTinterwordstretchfactor}{4}
\providecommand{\BIBentryALTinterwordspacing}{\spaceskip=\fontdimen2\font plus
\BIBentryALTinterwordstretchfactor\fontdimen3\font minus
  \fontdimen4\font\relax}
\providecommand{\BIBforeignlanguage}[2]{{%
\expandafter\ifx\csname l@#1\endcsname\relax
\typeout{** WARNING: IEEEtran.bst: No hyphenation pattern has been}%
\typeout{** loaded for the language `#1'. Using the pattern for}%
\typeout{** the default language instead.}%
\else
\language=\csname l@#1\endcsname
\fi
#2}}
\providecommand{\BIBdecl}{\relax}
\BIBdecl

\bibitem{vijayvargiya2025interactive}
S.~Vijayvargiya, X.~Zhou, A.~Yerukola, M.~Sap, and G.~Neubig, ``Interactive
  agents to overcome ambiguity in software engineering,'' \emph{arXiv preprint
  arXiv:2502.13069}, 2025.

\bibitem{mu2023clarifygpt}
F.~Mu, L.~Shi, S.~Wang, Z.~Yu, B.~Zhang, C.~Wang, S.~Liu, and Q.~Wang,
  ``Clarifygpt: Empowering llm-based code generation with intention
  clarification,'' \emph{arXiv preprint arXiv:2310.10996}, 2023.

\bibitem{le2014flashextract}
V.~Le and S.~Gulwani, ``Flashextract: A framework for data extraction by
  examples,'' in \emph{Proceedings of the 35th ACM SIGPLAN Conference on
  Programming Language Design and Implementation}, 2014, pp. 542--553.

\bibitem{kuhn2023semantic}
L.~Kuhn, Y.~Gal, and S.~Farquhar, ``Semantic uncertainty: Linguistic
  invariances for uncertainty estimation in natural language generation,''
  \emph{arXiv preprint arXiv:2302.09664}, 2023.

\bibitem{le2019automated}
C.~Le~Goues, M.~Pradel, and A.~Roychoudhury, ``Automated program repair,''
  \emph{Communications of the ACM}, vol.~62, no.~12, pp. 56--65, 2019.

\bibitem{fan2024oracle}
Z.~Fan, H.~Ruan, S.~Mechtaev, and A.~Roychoudhury, ``Oracle-guided program
  selection from large language models,'' in \emph{Proceedings of the 33rd ACM
  SIGSOFT International Symposium on Software Testing and Analysis}, 2024, pp.
  628--640.

\bibitem{liu2024deepseek}
A.~Liu, B.~Feng, B.~Xue, B.~Wang, B.~Wu, C.~Lu, C.~Zhao, C.~Deng, C.~Zhang,
  C.~Ruan \emph{et~al.}, ``Deepseek-v3 technical report,'' \emph{arXiv preprint
  arXiv:2412.19437}, 2024.

\bibitem{tian2025fixing}
Z.~Tian, J.~Chen, and X.~Zhang, ``Fixing large language models' specification
  misunderstanding for better code generation,'' in \emph{2025 IEEE/ACM 47th
  International Conference on Software Engineering (ICSE)}.\hskip 1em plus
  0.5em minus 0.4em\relax IEEE Computer Society, 2025, pp. 645--645.

\bibitem{fan2023automated}
Z.~Fan, X.~Gao, M.~Mirchev, A.~Roychoudhury, and S.~H. Tan, ``Automated repair
  of programs from large language models,'' in \emph{2023 IEEE/ACM 45th
  International Conference on Software Engineering (ICSE)}.\hskip 1em plus
  0.5em minus 0.4em\relax IEEE, 2023, pp. 1469--1481.

\bibitem{srivastava2013template}
S.~Srivastava, S.~Gulwani, and J.~S. Foster, ``Template-based program
  verification and program synthesis,'' \emph{International Journal on Software
  Tools for Technology Transfer}, vol.~15, pp. 497--518, 2013.

\bibitem{gleich2010ambiguity}
B.~Gleich, O.~Creighton, and L.~Kof, ``Ambiguity detection: Towards a tool
  explaining ambiguity sources,'' in \emph{Requirements Engineering: Foundation
  for Software Quality: 16th International Working Conference, REFSQ 2010,
  Essen, Germany, June 30--July 2, 2010. Proceedings 16}.\hskip 1em plus 0.5em
  minus 0.4em\relax Springer, 2010, pp. 218--232.

\bibitem{hou2023decomposing}
B.~Hou, Y.~Liu, K.~Qian, J.~Andreas, S.~Chang, and Y.~Zhang, ``Decomposing
  uncertainty for large language models through input clarification
  ensembling,'' \emph{arXiv preprint arXiv:2311.08718}, 2023.

\bibitem{petsios2017nezha}
T.~Petsios, A.~Tang, S.~Stolfo, A.~D. Keromytis, and S.~Jana, ``Nezha:
  Efficient domain-independent differential testing,'' in \emph{2017 IEEE
  Symposium on security and privacy (SP)}.\hskip 1em plus 0.5em minus
  0.4em\relax IEEE, 2017, pp. 615--632.

\bibitem{sharma2025assessing}
A.~Sharma and C.~David, ``Assessing correctness in llm-based code generation
  via uncertainty estimation,'' \emph{arXiv preprint arXiv:2502.11620}, 2025.

\bibitem{chen2022codet}
B.~Chen, F.~Zhang, A.~Nguyen, D.~Zan, Z.~Lin, J.-G. Lou, and W.~Chen, ``Codet:
  Code generation with generated tests,'' \emph{arXiv preprint
  arXiv:2207.10397}, 2022.

\bibitem{ding2024cycle}
Y.~Ding, M.~J. Min, G.~Kaiser, and B.~Ray, ``Cycle: Learning to self-refine the
  code generation,'' \emph{Proceedings of the ACM on Programming Languages},
  vol.~8, no. OOPSLA1, pp. 392--418, 2024.

\bibitem{hurst2024gpt}
A.~Hurst, A.~Lerer, A.~P. Goucher, A.~Perelman, A.~Ramesh, A.~Clark, A.~Ostrow,
  A.~Welihinda, A.~Hayes, A.~Radford \emph{et~al.}, ``Gpt-4o system card,''
  \emph{arXiv preprint arXiv:2410.21276}, 2024.

\bibitem{yang2024qwen2}
A.~Yang, B.~Yang, B.~Zhang, B.~Hui, B.~Zheng, B.~Yu, C.~Li, D.~Liu, F.~Huang,
  H.~Wei \emph{et~al.}, ``Qwen2.5 technical report,'' \emph{arXiv preprint
  arXiv:2412.15115}, 2024.

\bibitem{chen2021evaluating}
M.~Chen, J.~Tworek, H.~Jun, Q.~Yuan, H.~P. D.~O. Pinto, J.~Kaplan, H.~Edwards,
  Y.~Burda, N.~Joseph, G.~Brockman \emph{et~al.}, ``Evaluating large language
  models trained on code,'' \emph{arXiv preprint arXiv:2107.03374}, 2021.

\bibitem{austin2021program}
J.~Austin, A.~Odena, M.~Nye, M.~Bosma, H.~Michalewski, D.~Dohan, E.~Jiang,
  C.~Cai, M.~Terry, Q.~Le \emph{et~al.}, ``Program synthesis with large
  language models,'' \emph{arXiv preprint arXiv:2108.07732}, 2021.

\bibitem{liu2023your}
J.~Liu, C.~S. Xia, Y.~Wang, and L.~Zhang, ``Is your code generated by chatgpt
  really correct? rigorous evaluation of large language models for code
  generation,'' \emph{Advances in Neural Information Processing Systems},
  vol.~36, pp. 21\,558--21\,572, 2023.

\bibitem{jain2025livecodebench}
N.~Jain, K.~Han, A.~Gu, W.-D. Li, F.~Yan, T.~Zhang, S.~Wang, A.~Solar-Lezama,
  K.~Sen, and I.~Stoica, ``Livecodebench: Holistic and contamination free
  evaluation of large language models for code,'' in \emph{The Thirteenth
  International Conference on Learning Representations}, 2025.

\bibitem{li2022competition}
Y.~Li, D.~Choi, J.~Chung, N.~Kushman, J.~Schrittwieser, R.~Leblond, T.~Eccles,
  J.~Keeling, F.~Gimeno, A.~Dal~Lago \emph{et~al.}, ``Competition-level code
  generation with alphacode,'' \emph{Science}, vol. 378, no. 6624, pp.
  1092--1097, 2022.

\bibitem{xia2024top}
C.~S. Xia, Y.~Deng, and L.~Zhang, ``Top leaderboard ranking= top coding
  proficiency, always? evoeval: Evolving coding benchmarks via llm,''
  \emph{arXiv preprint arXiv:2403.19114}, 2024.

\bibitem{monperrus2014critical}
M.~Monperrus, ``A critical review of" automatic patch generation learned from
  human-written patches": Essay on the problem statement and the evaluation of
  automatic software repair,'' in \emph{Proceedings of the 36th International
  Conference on Software Engineering}, 2014, pp. 234--242.

\bibitem{iglewicz1993volume}
B.~Iglewicz and D.~C. Hoaglin, \emph{Volume 16: how to detect and handle
  outliers}.\hskip 1em plus 0.5em minus 0.4em\relax Quality Press, 1993.

\bibitem{uddin2020mining}
G.~Uddin, F.~Khomh, and C.~K. Roy, ``Mining api usage scenarios from stack
  overflow,'' \emph{Information and Software Technology}, vol. 122, p. 106277,
  2020.

\bibitem{hirsch2021identifying}
T.~Hirsch and B.~Hofer, ``Identifying non-natural language artifacts in bug
  reports,'' in \emph{2021 36th IEEE/ACM International Conference on Automated
  Software Engineering Workshops (ASEW)}.\hskip 1em plus 0.5em minus
  0.4em\relax IEEE, 2021, pp. 191--197.

\bibitem{fan2024self}
L.~Fan, M.~Chen, and Z.~Liu, ``Self-explained keywords empower large language
  models for code generation,'' \emph{arXiv preprint arXiv:2410.15966}, 2024.

\bibitem{li2024generating}
J.~Li, G.~Li, L.~Wang, H.~Zhu, and Z.~Jin, ``Generating equivalent
  representations of code by a self-reflection approach,'' \emph{arXiv preprint
  arXiv:2410.03351}, 2024.

\bibitem{liu2024codemind}
C.~Liu, S.~D. Zhang, A.~R. Ibrahimzada, and R.~Jabbarvand, ``Codemind: A
  framework to challenge large language models for code reasoning,''
  \emph{arXiv preprint arXiv:2402.09664}, 2024.

\bibitem{ezzini2022automated}
S.~Ezzini, S.~Abualhaija, C.~Arora, and M.~Sabetzadeh, ``Automated handling of
  anaphoric ambiguity in requirements: a multi-solution study,'' in
  \emph{Proceedings of the 44th International Conference on Software
  Engineering}, 2022, pp. 187--199.

\bibitem{luitel2023using}
D.~Luitel, S.~Hassani, and M.~Sabetzadeh, ``Using language models for enhancing
  the completeness of natural-language requirements,'' in \emph{International
  working conference on requirements engineering: foundation for software
  quality}.\hskip 1em plus 0.5em minus 0.4em\relax Springer, 2023, pp. 87--104.

\bibitem{ferrari2019nlp}
A.~Ferrari and A.~Esuli, ``An nlp approach for cross-domain ambiguity detection
  in requirements engineering,'' \emph{Automated Software Engineering},
  vol.~26, no.~3, pp. 559--598, 2019.

\bibitem{ferrari2018detecting}
A.~Ferrari, G.~Gori, B.~Rosadini, I.~Trotta, S.~Bacherini, A.~Fantechi, and
  S.~Gnesi, ``Detecting requirements defects with nlp patterns: an industrial
  experience in the railway domain,'' \emph{Empirical Software Engineering},
  vol.~23, no.~6, pp. 3684--3733, 2018.

\bibitem{fantechi2023inconsistency}
A.~Fantechi, S.~Gnesi, L.~Passaro, and L.~Semini, ``Inconsistency detection in
  natural language requirements using chatgpt: a preliminary evaluation,'' in
  \emph{2023 IEEE 31st International Requirements Engineering Conference
  (RE)}.\hskip 1em plus 0.5em minus 0.4em\relax IEEE, 2023, pp. 335--340.

\bibitem{ma2024large}
R.~Ma, X.~Wang, X.~Zhou, J.~Li, N.~Du, T.~Gui, Q.~Zhang, and X.~Huang, ``Are
  large language models good prompt optimizers?'' \emph{arXiv preprint
  arXiv:2402.02101}, 2024.

\bibitem{ruan2024specrover}
H.~Ruan, Y.~Zhang, and A.~Roychoudhury, ``Specrover: Code intent extraction via
  llms,'' \emph{arXiv preprint arXiv:2408.02232}, 2024.

\bibitem{ma2024specgen}
L.~Ma, S.~Liu, Y.~Li, X.~Xie, and L.~Bu, ``Specgen: Automated generation of
  formal program specifications via large language models,'' \emph{arXiv
  preprint arXiv:2401.08807}, 2024.

\bibitem{endres2024can}
M.~Endres, S.~Fakhoury, S.~Chakraborty, and S.~K. Lahiri, ``Can large language
  models transform natural language intent into formal method postconditions?''
  \emph{Proceedings of the ACM on Software Engineering}, vol.~1, no. FSE, pp.
  1889--1912, 2024.

\bibitem{le2011genprog}
C.~Le~Goues, T.~Nguyen, S.~Forrest, and W.~Weimer, ``Genprog: A generic method
  for automatic software repair,'' \emph{Ieee transactions on software
  engineering}, vol.~38, no.~1, pp. 54--72, 2011.

\bibitem{mechtaev2016angelix}
S.~Mechtaev, J.~Yi, and A.~Roychoudhury, ``Angelix: Scalable multiline program
  patch synthesis via symbolic analysis,'' in \emph{Proceedings of the 38th
  international conference on software engineering}, 2016, pp. 691--701.

\bibitem{xia2023keep}
C.~S. Xia and L.~Zhang, ``Keep the conversation going: Fixing 162 out of 337
  bugs for \$0.42 each using chatgpt,'' \emph{arXiv preprint arXiv:2304.00385},
  2023.

\bibitem{bouzenia2024repairagent}
I.~Bouzenia, P.~Devanbu, and M.~Pradel, ``Repairagent: An autonomous, llm-based
  agent for program repair,'' \emph{arXiv preprint arXiv:2403.17134}, 2024.

\bibitem{parasaram2024fact}
N.~Parasaram, H.~Yan, B.~Yang, Z.~Flahy, A.~Qudsi, D.~Ziaber, E.~Barr, and
  S.~Mechtaev, ``The fact selection problem in llm-based program repair,''
  \emph{arXiv preprint arXiv:2404.05520}, 2024.

\end{thebibliography}


\end{document}